\documentclass[final,5p,times,twocolumn,authoryear]{elsarticle}
\usepackage{amssymb}
\usepackage{hyperref}
\usepackage{aas_macros}
\usepackage[nolist]{acronym}

\journal{Astronomy and Computing}

\begin{acronym}
\acro{MMA}{multi-messenger astrophysics}
\acro{SSC}{Science Support Center}
\acro{CAP}{community access portal}
\acro{EM}{electromagnetic}
\acro{GW}{gravitational wave}
\end{acronym}

\begin{document}

\begin{frontmatter}
\title{The NASA Multi-Messenger Astrophysics Science Support Center (MOSSAIC)*}

\author[gsfc]{Rita M. Sambruna\corref{corresponding-author}}\ead{rita.m.sambruna@nasa.gov}
\author[gsfc]{Joshua E. Schlieder\corref{corresponding-author}}\ead{joshua.e.schlieder@nasa.gov}
\author[msfc]{Daniel Kocevski\corref{corresponding-author}}\ead{daniel.kocevski@nasa.gov}

\author[gsfc]{Regina Caputo}
\author[msfc]{Michelle C. Hui}
\author[gsfc]{Craig B. Markwardt}
\author[gsfc]{Brian P. Powell}
\author[gsfc]{Judith L. Racusin}
\author[gsfc]{Christopher Roberts}
\author[gsfc]{Leo P. Singer}
\author[gsfc]{Alan P. Smale}
\author[gsfc]{Tonia M. Venters}
\author[msfc]{Colleen A. Wilson-Hodge}

\cortext[corresponding-author]{Corresponding author}

\affiliation[gsfc]{
    organization={NASA Goddard Space Flight Center},
    addressline={8800 Greenbelt Rd.}, 
    city={Greenbelt},
    postcode={20771}, 
    state={MD},
    country={USA}
}

\affiliation[msfc]{
    organization={NASA Marshall Space Flight Center},
    addressline={Martin Rd. SW}, 
    city={Huntsville},
    postcode={35808}, 
    state={AL},
    country={USA}}

\begin{abstract}
The era of \acl{MMA} has arrived, leading to key new discoveries and revealing a need for coordination, collaboration, and communication between world-wide communities using ground and space-based facilities. To fill these critical needs, NASA’s Goddard Space Flight Center and Marshall Space Flight Center are jointly proposing to establish a virtual Multi-Messenger Astrophysics Science Support Center that focuses entirely on community-directed services. In this article, we describe the baseline plan for the virtual Support Center which will position the community and NASA as an Agency to extract maximum science from multi-messenger events, leading to new breakthroughs and fostering increased coordination and collaboration. 

    \textit{Note: between the time of the paper submission and receipt of the referee report, the MMA SSC evolved into MOSSAIC - Multimessenger Operational Astrophysics and Information Collaboration, which besides GSFC and MSFC includes non-NASA institutions; see \url{http://asd.gsfc.nasa.gov/mossaic}}
\end{abstract}



\begin{keyword}
\href{http://astrothesaurus.org/uat/78}{Astroinformatics} \sep
\href{http://astrothesaurus.org/uat/1856}{Astronomy web services} \sep
\href{http://astrothesaurus.org/uat/739}{High energy astrophysics} \sep
\href{http://astrothesaurus.org/uat/678}{Gravitational waves} \sep
\href{http://astrothesaurus.org/uat/1100}{Neutrino astronomy} \sep
\href{http://astrothesaurus.org/uat/324}{Cosmic ray astronomy}
\end{keyword}

\end{frontmatter}

\section{Introduction}
\label{sec-intro}

\Ac{MMA} has come of age thanks 
to the detection of \ac{GW} sources from 
the ground with the Advanced LIGO \citep{2015CQGra..32g4001L} and Virgo \citep{2015CQGra..32b4001A} observatories, and of 
an extra-galactic neutrino source with the IceCube Neutrino
Observatory \citep{2017JInst..12P3012A}. Together with the concurrent observations of 
coincident gamma-ray photons followed by photons at other
\ac{EM} wavelengths, these discoveries 
provide new insights into the physics of the Universe. 
While at this time the 2020 Astrophysics Decadal Survey report is still to be 
released, it is expected that strong recommendations will 
be made for \ac{MMA}.  

The advent of advanced ground-based observatories in a few
years will expand the discovery horizon and drastically 
increase  the number of sources needing prompt  EM 
follow-up from the ground and in space. The needs of the 
\ac{MMA} community will increase many-fold. This includes  the 
need for coordination, collaboration, and communication 
(the 3Cs) between space and ground-based facilities; the 
need for adequate infrastructure---data analysis and 
interpretation tools, efficient alert systems, proposer 
and observer support, rapid data transmission links, etc.;
and the need for common and frequent transfer of ideas 
between communities to anticipate future needs and provide
solutions. A similar conclusion was previously reached by \citep{2019ASPC..523..503K}, where they made specific suggestions for new communication protocols using VO. 

To fill these critical needs, NASA’s Goddard Space Flight 
Center (GSFC) and Marshall Space Flight Center (MSFC) are 
jointly proposing to establish a virtual MMA \ac{SSC}, with 100\% community-directed 
services. Here we describe the baseline plan for the 
virtual MMA SSC which will provide:  

\begin{figure*}[t!]
\centering \includegraphics[width=19cm]{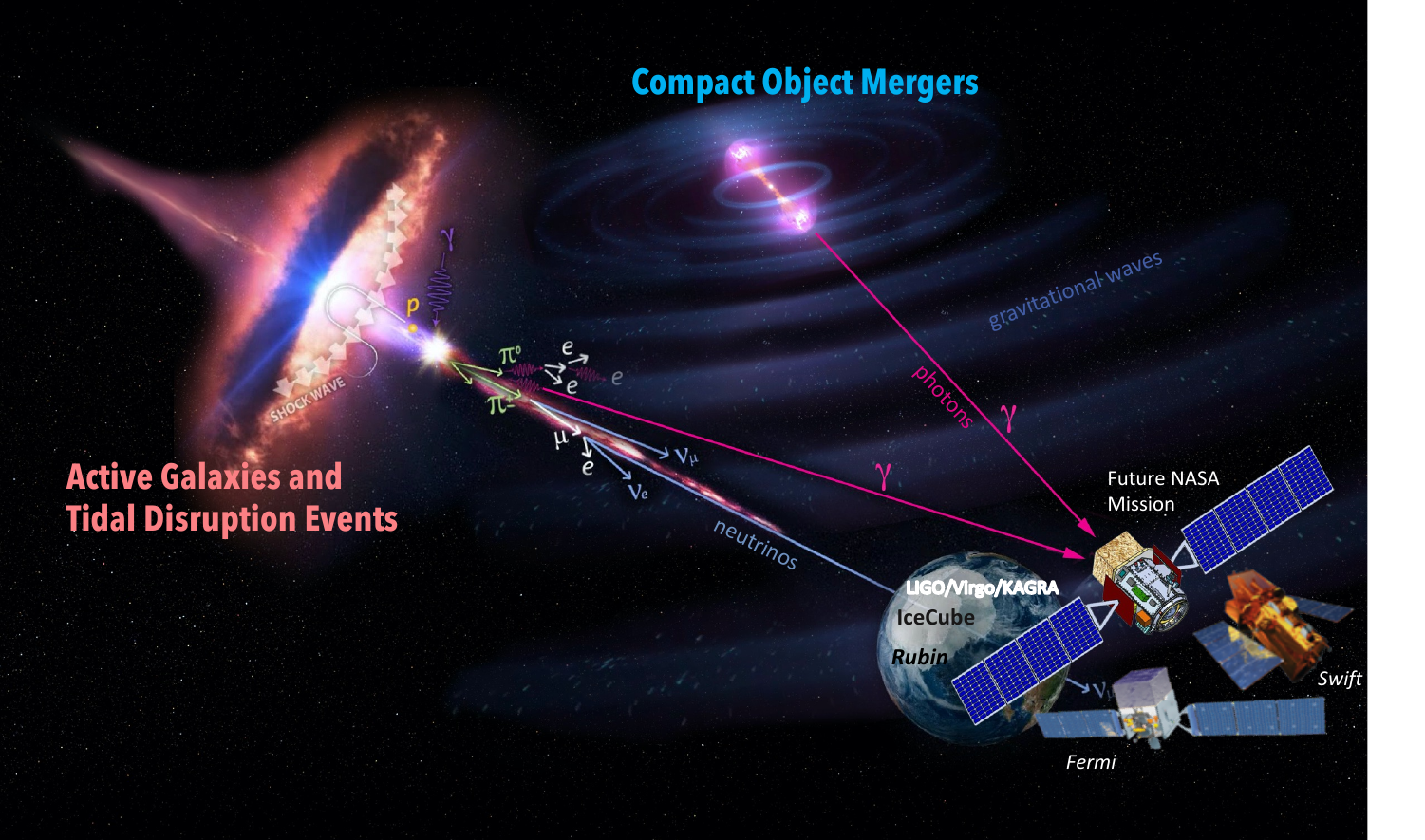}
\caption{\Ac{MMA} focuses on extreme astronomical events that produce multiple types of messengers, gravitational waves, neutrinos, cosmic rays, and photons, that each convey key information about the event and the underlying physics. A complete understanding of these events requires coordination, collaboration, and communication between multiple communities and observatories on the ground and in space. \textit{Schematic courtesy of the AMEGO/AMEGO-X \citep{2019BAAS...51g.245M} mission concept teams, adapted by J. Schlieder.}}
\label{fig-mma_schematic}
\end{figure*}

\begin{itemize}
    \item {A \ac{CAP} website to act as a one-stop-shop for information, tools, and support.}
    \item {A real-time board for instant community collaboration, coordination, and communication and dissemination of observing plans.}
    \item {A modern alert system for NASA missions and other current Gamma-ray Coordinates Network/Transient Astronomy Network (GCN/TAN\footnote{\url{https://gcn.gsfc.nasa.gov/}}) streams, built on technologies embraced by the \ac{MMA} community (e.g., Apache Kafka).}
    \item{A Guest Observer Facility (GOF)-like service for community support in observing and proposing to NASA and ground based facilities for \ac{MMA} targets.}
    \item{A suite of tools for analysis and interpretation of data from NASA missions.}
    \item{A curated archive of data serving the specific needs of the \ac{MMA} community and the development of relevant analysis tools, including automation of certain functions and analysis through artificial intelligence/machine learning (AI/ML).}
    \item{A service where new and existing MMA community members can obtain consultation, tools, and expertise to improve MMA cross-integration and to design new missions.}
    \item{Expertise and experience in science definition for MMA missions led by external Principal Investigators (PIs).}
    \item{Community building and networking events (workshops, conferences, training), with a special focus on recruiting and retaining a diverse workforce at the NASA Centers, and bringing together all communities involved in MMA science.}
\end{itemize} 

NASA’s MMA SSC will make it possible for the community to 
reap maximum benefit from MMA science and missions, 
providing coordination and facilitating collaboration. MMA
is by definition a team enterprise, and the 3Cs---collaboration, coordination, and communication---are at 
the heart of the MMA SSC. We acknowledge and support 
ongoing independent efforts for MMA in the  scientific 
community; our aim is to connect with them and amplify 
their services and impact, not replace them. We invite the
broader community to contact us with additional ideas for 
collaboration.  

\section{Multi-Messenger Astrophysics in Context}
\label{sec-context}

\begin{figure}[htb!]
\includegraphics[width=\linewidth]{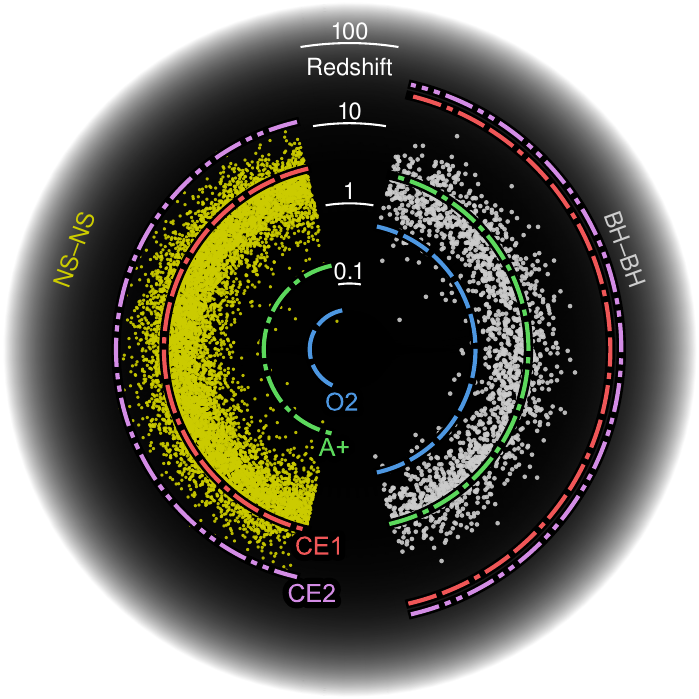}
\caption{The advent of improved sensitivities for GW detection on the ground will expand the horizon of the GW universe in need of EM follow-up in space and from the ground. NS-NS = neutron star - neutron star merger; BH-BH = Black hole - black hole merger; O2 = LIGO/Virgo Observing Run 2; A+ = Advanced LIGO Plus; CE1, CE2 = Cosmic Explorer (two different sensitivities). \textit{Reproduced from \citet{2019BAAS...51c.141R}.}}
\label{fig-horizon}
\end{figure}

\Ac{MMA} has come of age thanks 
to the detection of \ac{GW} sources with 
the ground-based LIGO and Virgo observatories, and of an 
extragalactic neutrino source with the ground-based 
IceCube Neutrino Observatory. Together with the concurrent
observations of high-energy photons, these discoveries 
provided new insights into the physics of the Universe, 
with rippling consequences for other science disciplines 
as well (e.g., chemistry, fundamental physics, etc.). The 
first joint GW and \ac{EM} detection of a 
binary neutron star merger (GW170817) by the Fermi 
Gamma-ray Burst Monitor \citep[GBM][]{2009ApJ...702..791M} and by ESA’s INTEGRAL 
mission \citep{2003A&A...411L...1W} revolutionized our knowledge of these systems \citep{2017ApJ...848L..12A}.  In the four years since its 
detection, over 4000 papers have cited the GW170817 
discovery paper, on topics ranging from nuclear physics to
radiation transport, general relativity, and relativistic 
astrophysics. Likewise, the recent detection of a 
high-energy neutrino (IC170922) correlated in space and 
time with a flare from gamma-ray blazar TXS 0506+056 \citep[LAT;][]{2018Sci...361..147I,2018Sci...361.1378I}
detected by the Fermi Large Area Telescope \citep[LAT][]{2012ApJS..203....4A}, and the 
possible association of a high-energy neutrino with a  
tidal disruption event \citep[TDE][]{2021NatAs...5..510S}, 
has provided a tantalizing clue to the origin of high-energy 
cosmic neutrinos. In the coming years, the advent of A+ 
LIGO/Virgo/KAGRA/LIGO-India will catapult the detection 
rate of GW sources to several per month or even per week, 
placing increased strain on the search for their EM 
counterparts from the ground and in space (see Fig.~\ref{fig-horizon}). IceCube-Gen2 \citep{2021JPhG...48f0501A} 
will similarly increase the number of neutrino detections 
that require EM counterpart follow-up. 

Identification and characterization of the EM counterparts of GW/neutrino sources is central to fulfilling the promise of MMA. Many challenges require progress \citep{2019BAAS...51c.123B,2019BAAS...51g.154M,2019BAAS...51g.139M}: poor GW localizations, implying thousands of possible optical counterparts and posing the issue of how to identify the right one effectively and efficiently; coordinating and optimizing the observing windows of a transient from space and ground-based telescopes, taking into account the different observational constraints of the various observatories; improving alert systems (e.g., upgrading the GCN and/or migrating to new systems) and related dissemination pipelines; restructuring the data archives and related software for optimal MMA science; and even optimizing communication protocols for faster responses (space to ground, space to space, and ground to ground) with increased cybersecurity. Given the volume of data and the necessary management, some degree of automation in analysis tools is desirable as well.  

Moreover, the study of the time-variable Universe will soon undergo a revolution with the advent of the ground-based Vera Rubin Observatory (Rubin) and its Large Survey of Space and Time \citep[LSST;][]{2019ApJ...873..111I}, expected to detect millions of transients per night, augmenting the already operating Zwicky Transient Factory \citep[ZTF;][]{2019PASP..131a8002B,2019PASP..131g8001G,2020PASP..132c8001D,2019PASP..131a8003M} and other current and planned facilities. (Note: while MMA is distinct from general time-domain astronomy, the latter also involves EM follow-up, so they will be considered together in this document under the overall term MMA.) It is clear that the MMA community needs to organize itself if we are to take maximum advantage of the upcoming and already operating facilities.  Several groups are already active (e.g., Scalable Infrastructure to support MMA, or SCiMMA\footnote{\url{https://scimma.org/}}, and Astrophysical Multi-messenger Observatory Network, or AMON \citep{2013APh....45...56S} and actively implementing/developing approaches to particular aspects of the data management (mainly from ground-based telescopes), the alerts system (for ZTF/Rubin), and individual campaign coordination (e.g., the past NSF-funded GROWTH\footnote{\url{https://www.growth.caltech.edu/}}). However, a coherent, cohesive framework bringing together ground and space-based communities is so far lacking, as well as a focused strategy for how to optimize the MMA science promise in the era of next generation GW and neutrino observatories. More importantly, NASA is currently lacking a modern alert system which meets the community demands of rapid turnaround for EM facilities in space. Similarly, faster and more efficient data communication is needed.  

NASA Centers are uniquely positioned to take the lead in coordinating NASA’s response to the MMA challenges of the 21st century. First, many NASA scientists have been at the forefront of MMA research, both observational and theoretical, and deeply involved in currently operating and recently selected missions, as well as mission concept studies; this means a wealth of knowledge and experience on MMA issues, which will only grow in the years ahead. Second, existing specific capabilities (High Energy Astrophysics Science Archive Research Center (HEASARC), GCN, LIGO-Virgo localization procedures) and expertise (e.g., AI/ML, GOF experience, mission operations, rapid multiwavelength follow-up, e.g., \textit{Swift} \citep{2004ApJ...611.1005G}, \textit{Fermi}, NICER \citep{2012SPIE.8443E..13G}, GBM pipeline, data analysis, and subthreshold searches for GRBs) make the NASA Centers the obvious hub for setting up an MMA nexus. Finally, NASA scientists have excellent professional relationships with the various MMA communities (both ground and space-based) ensuring direct coordination, support to and from, and no duplication of effort in the community. While the 2020 Astrophysics Decadal Survey is expected to make strong recommendations for enhancing MMA, both in terms of missions and community research infrastructure (see below), we can proactively start building a community focused MMA \ac{SSC} distributed across GSFC and MSFC.   

\section{Overview of the MMA SSC}
\label{sec-overview}

We are proposing to establish an MMA SSC at NASA, with the following aims:  

\begin{enumerate}
    \item {Provide resources and tools for MMA science to the astrophysics community to maximize the science return from individual events and population studies.}
    \item{Connect and promote collaborations among the MMA focused science community, including individuals, teams, and academic institutions.} 
\end{enumerate}

The MMA SSC leverages important existing MMA activities at Goddard and Marshall, led by our scientists with internal and external support. These activities, described in the next section, provide the pillars of the SSC, whose vision is to augment and extend their functionalities as a broader community service component.  

Science is a collaborative effort, and this is particularly true of MMA which lies at the intersection of many sub-fields: ground and space-based observation, computation, data manipulation and communication, innovative technology, IT infrastructure, and more. Thus, a key part of the proposal is our interdisciplinary collaboration with other NASA Goddard Directorates (including Information Technology and Space Navigation and Communication), and with academic and other institutions. Our collaborators provide key components of the SSC, leveraging their unique strengths. At the same time, the various pieces are independent of each other, which allows self-paced progress and decreases the overall project risk.   

\begin{figure}[htb!]
\includegraphics[width=\linewidth]{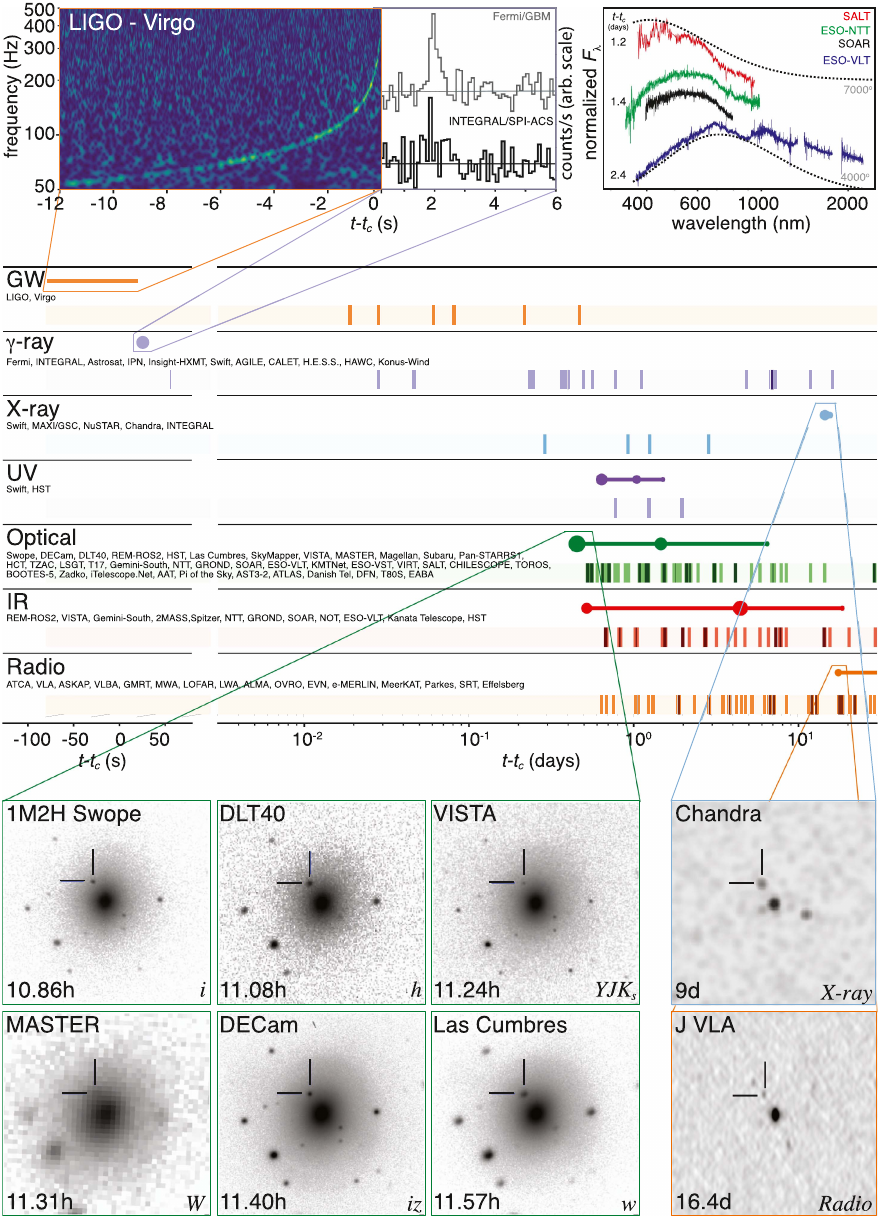} \\\\
\includegraphics[width=\linewidth]{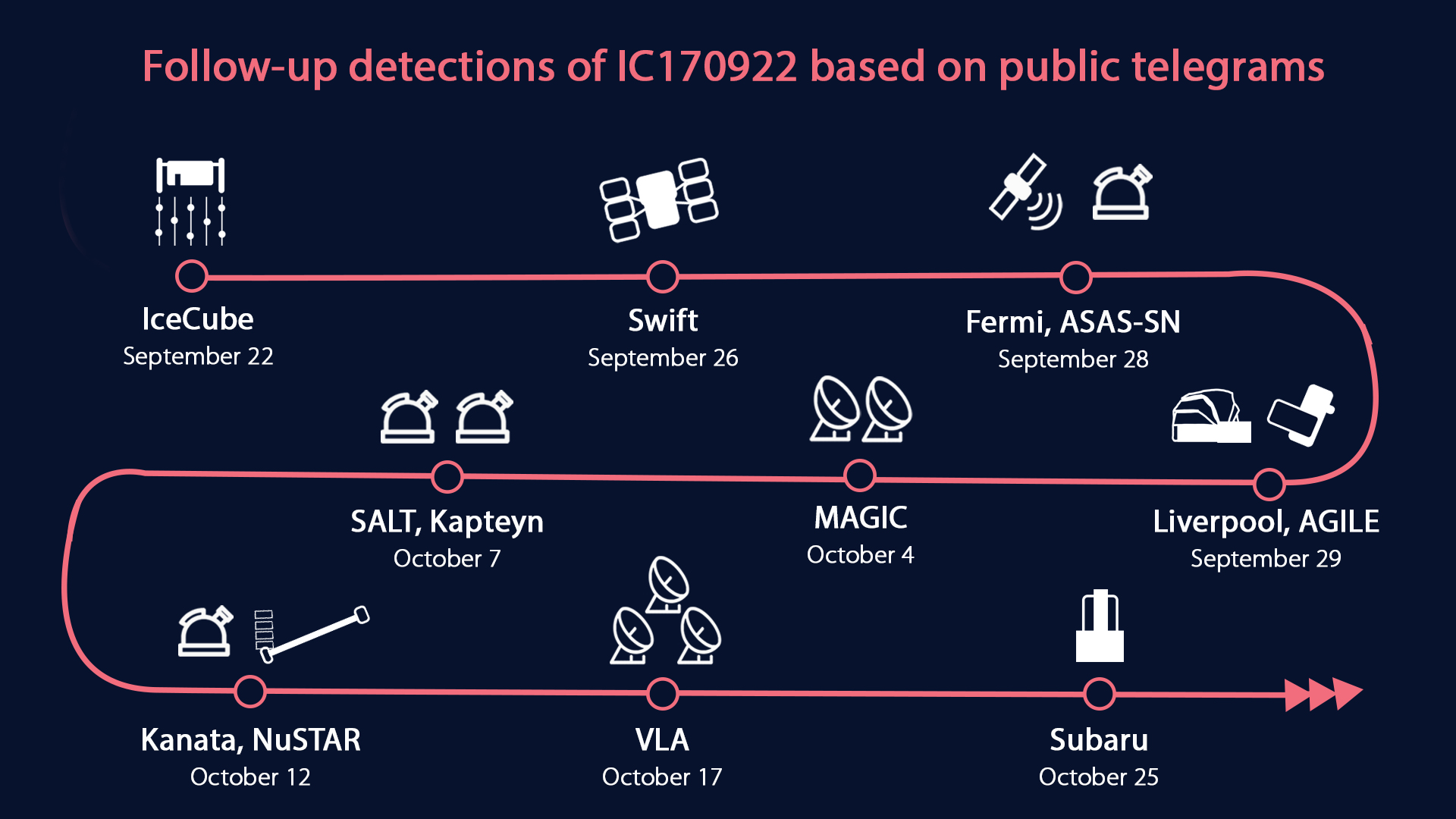}
\caption{Timescale of the follow-up campaign for the gravitational wave source GW170817 \citep[top panel;][]{2017ApJ...848L..12A} and the neutrino source IC170922 \citep[bottom panel;][]{2018Sci...361.1378I}, illustrating the level of coordination and labor required for each of these individual events. The rate of detection will undoubtedly increase significantly with the advent of the next generation of detectors, requiring a new level of coordination and planning. The MMA SSC will facilitate the necessary collaboration, coordination, and communication. \textit{Top panel reproduced from \citet{2017ApJ...848L..12A}. Bottom panel reproduced with permission from the IceCube Collaboration.}}
\label{fig-timelines}
\end{figure}

\begin{figure*}[htb!]
\centering \includegraphics[width=\textwidth]{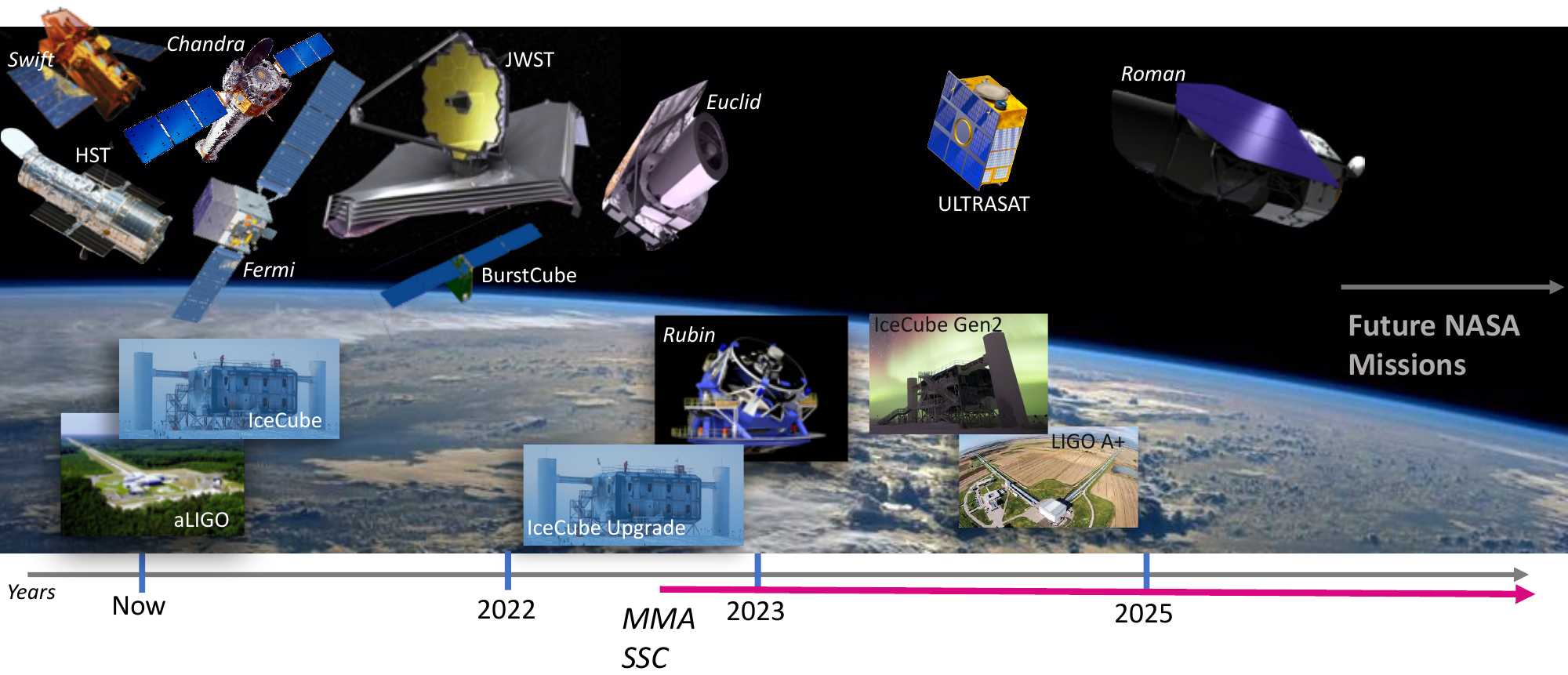}
\caption{The MMA SSC is timely when placed in context with current and future advanced facilities in space and on the ground. The MMA SSC would begin service to the community around the same time as MMA capable facilities are reaching their full potential (pink horizontal arrow).} \label{fig-mmassc_timelines}
\end{figure*}

\textbf{The MMA SSC fills a critical gap in the astrophysics landscape of the 2020s and beyond.}  The landscape of MMA in the 2020s and beyond will be very different---and much more complex---than at the times of the momentous MMA events of the current era: the GW170817 and IC170922A detections. The advanced sensitivity of ground-based GW and neutrino detectors will allow us to probe a much greater volume of the local Universe (out to $>$300 Mpc for A+ LIGO), implying that events like GW170817 could be detected monthly, or even more frequently. For example, estimates for the future rate of binary neutron star merger detections alone are in the range 10--200\,yr$^{-1}$ \citep{2020LRR....23....4B,2019BAAS...51c.141R}.  Keeping up with this rate will require a much higher degree of cooperation and coordination among the follow-up facilities, as well as strategic planning in advance of  specific alerts. Figure~\ref{fig-mmassc_timelines} shows the timeline for the MMA SSC in context.

\textbf{A NASA-supported MMA SSC provides meaningful impacts for NASA and the space- and ground-based communities.} By acting as the bridge between various MMA communities, and by bringing together observers and theorists to plan ahead and coordinate for optimal strategies, the MMA SSC fulfills a critical need for MMA science. In this sense, the MMA SSC is not just desirable \textit{but essential}; with a modest investment of funds, NASA will: 1. Leverage the science return on NASA missions; 2. Maintain international leadership, expertise, and capabilities in MMA and time-domain astronomy; and 3. Multiply its impact on the community by orders of magnitude (as shown by GW170817).

Advance planning will be key, including lining up observatories to be ready to go as soon as alerts are provided, to narrow in on the time of the event and the few moments after, which are critical to our understanding of the physics. Therefore, while in 2017 the community self-organized around single events with remarkable success, this will be impossible going forward due to the sheer number of events. In addition, the Vera Rubin Observatory will provide millions of transients (not all of MMA interest) per night of observation, raising the issue of how to select targets for NASA follow-ups, and with which observatories. Not to mention the need for a robust, flexible, and more reliable alert dissemination system matched to the rate of discoveries. And finally, providing a suite of analysis and interpretation tools will level the playing field for the entire community, enabling scientists from small colleges and Minority Serving Institutions (MSIs)  to contribute to MMA science.

\section{MMA Community Support at NASA Centers}
\label{sec-support}

The role of NASA’s missions in MMA can not be overemphasized, as operating in space is the only way to access the higher energies of the \ac{EM} spectrum, where the extreme physics of MMA sources lead to strong emission. Undoubtedly, the Neil Gehrels Swift Observatory and Fermi have been the workhorses of most of the EM follow-ups of \ac{GW} sources at high energies; Swift also provides access to UV wavelengths for important early kilonova searches. In addition, NASA Centers have provided and will continue to provide critical support to the MMA community via their services, activities, and expertise in data analysis, archiving and computing, space communications, and mission concept development. 

Goddard and Marshall are leaders in service to the MMA community and in facilitating broad participation in MMA. The HEASARC, hosted at GSFC, is NASA’s archive for high-energy astrophysics (HEA) and cosmic microwave background (CMB) data. The HEASARC’s innovative archive interfaces and multi-mission software allow scientists to identify, download, display, correlate, and analyze scientific data from dozens of past and current missions, including those led by other Agencies outside the US (e.g., JAXA and ESA). 

The GCN/TAN  service was integrated permanently into the HEASARC in October 2016. As a result the GCN hardware was upgraded in 2017, and a large number of new notice types have recently been added. Goddard and the GCN already collaborate closely with the LIGO-Virgo consortium, AMON, and IceCube to disseminate multimessenger alerts for \ac{GW} events, high energy neutrino detections, and other non-EM-spectrum-based alerts. GCN is central to the Time-domain Astronomy Coordination Hub (TACH) multimessenger activity, and work is already under way to expand and modernize GCN and to provide new interfaces and tools to the alerts. 

In collaboration with Goddard’s Near Space Network, a new low-latency space communications service access protocol was deployed by the Neil Gehrels Swift Observatory in 2020. The protocol has enabled science’s first known fully autonomous spacecraft-commanding data pipeline, dramatically improving follow-up response time and scientific yields for transients detected by other instruments \citep{2020ApJ...900...35T}. The Near Space Network has invested in continued research and development to address the challenges MMA observation concepts pose to space communications network planning, design, and operations.

Goddard and Marshall also support the MMA community and enable MMA science through various other outlets. Goddard leads the Guest Observer/Guest Investigator (GO/GI) programs for NASA’s Explorer class missions.  Goddard scientists develop GO/GI proposal calls, administer GO/GI reviews, and manage and maintain mission Science Support Centers (SSCs). SSCs provide the community with all relevant mission information and updates through primary websites, develop and maintain community software tools, and operate helpdesks to field questions from the community. SSCs also represent missions at national and international conferences and workshops, providing a direct interface between the science community and the missions. Marshall scientists have likewise developed modern open-source data reduction toolkits that allow scientists to simultaneously analyze data from multiple gamma-ray instruments that observe the same transient and have led the effort to produce automated joint MMA localizations.

The proposed MMA SSC will be physically hosted at GSFC with virtual participation from the other Centers. Internal GSFC resources have already been committed to kickstart the MMA SSC effort. As discussed above, the anticipated activities of the MMA SSC will be 100\% community service oriented and aiming to leverage, not duplicate, existing capabilities. In the following section, we describe the proposed activities in some detail.

\section{Functions of the NASA MMA SSC}
\label{sec-functions}

The primary goal of the NASA MMA SSC is to foster excellent MMA science through 100\% community service, building on the expertise and capabilities of Goddard and Marshall. Specifically, the proposed tasks of the MMA SSC include:

\begin{itemize}
    \item A \ac{CAP} website to act as a one stop shop for information, tools, and support. The CAP website will be the port of entry for MMA community members seeking the services of the MMA SSC. We will provide clear and direct access to all relevant information needed for MMA community members to facilitate their science. The front page of the CAP will include general MMA SSC information and regular news updates. The front page will also include a sidebar with key dates (e.g., upcoming conferences and proposal deadlines), access to the Helpdesk (see below), and feeds of MMA related social media (from e.g. NASA, NSF, Rubin, etc.). The front page will also provide links to all of the other core functions of the community portal as pull down menus:
    \begin{itemize}
        \item a Helpdesk, to address open community questions related to MMA science and missions;
        \item tools for MMA science, widely varying and in the form of web based user interfaces, downloadable software packages, and useful data sets that will be provided by both the MMA SSC and by the community; 
        \item Updated and modernized community tools to coordinate, disseminate, and archive rapid alerts related to MMA sources and other transients; 
        \item A real-time board for instant community communication and dissemination of observing plans, where various observers can alert one another about who is following which event with what facility;
        \item A description of MMA missions in development; and a board of events for the community (workshops, conferences, job opportunities); 
    \end{itemize}
    \item A modern alert system for NASA missions and other GCN streams, based upon the Apache Kafka technology being utilized by the ground-based optical transient community (e.g., Rubin). Once in place, it can be expanded upon for connection to additional communities, transient source classes, and other existing notification systems (e.g.,  SCiMMA, Rubin/LSST Brokers, etc.); 
    \item A Guest Observer Facility-like service for community support in observing and proposing. This will include:
    \begin{itemize}
        \item A Target Visibility Planner, a one-stop-shop that contains visibility and field-of-view constraints of many space-based observatories, as well as their current and predicted orbital tracks, which can be used to plan target visibility for multiple observatories (ground and space-based), visualize fields of view, and more. Our Target Visibility Planner will include scheduling constraints from missions both in space and on the ground, optimizing use of facilities in both domains and leveraging on existing tools \citep[e.g., ObsLocTAP][]{2021ivoa.spec.0724S}; 
        \item A Follow-Up Hub service, to help observers rapidly request and schedule follow-up observations of a target; and other GOF-like services providing tools for analysis and interpretation of MMA data. 
        \item Direct coordination with ground based facilities for optimal use of capabilities.
    \end{itemize}
    \item A curated archive of data serving the specific needs of the MMA community and the development of relevant analysis tools, including possible automation of certain functions and analysis through AI/ML. 
    \item A support group from which new and existing MMA community members can obtain consultation, tools, and expertise to improve MMA cross-integration and to evaluate new mission concepts.  This includes support for building the science case for new missions which builds on the computational capabilities and programmatic experience of our scientists.
    \item Organization of community-building and networking events (workshops, conferences, training), with a particular emphasis on training the diverse workforce of tomorrow.  We aim at bringing together not just the observers but also the infrastructure scientists, since planning and developing infrastructure (software, tools) is informed by the science needs. 
\end{itemize}

The MMA SSC will serve as the focal point for Goddard’s space communications and navigation community to engage and collaborate on MMA science concepts, understand emerging network service requirements, formulate network solutions, and align the space communications and navigation technology roadmap with the needs of the MMA science community.

\begin{figure}
    \centering
    \includegraphics[width=\columnwidth]{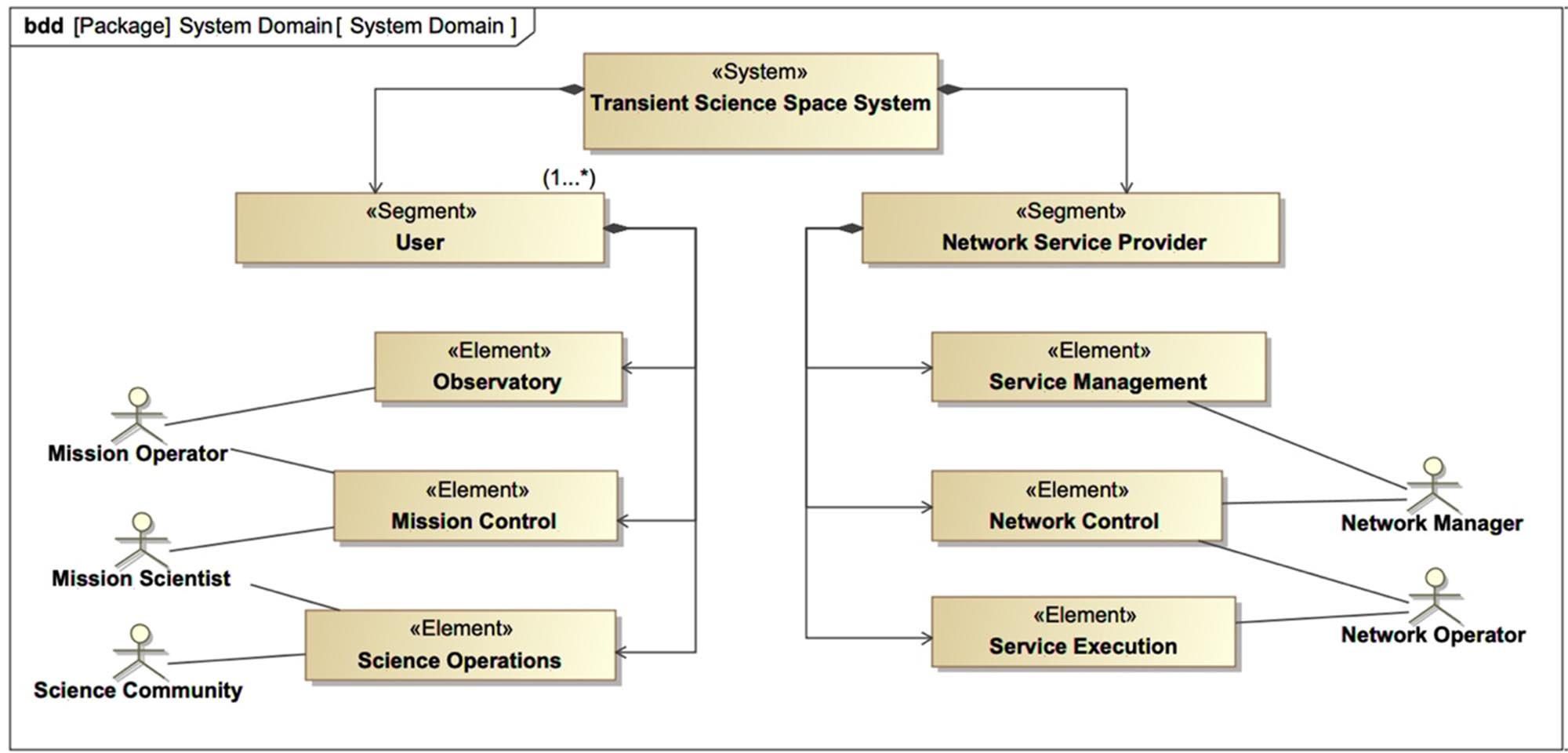}
    \caption{Top-level structural decomposition of an MMA transient science space system using Systems Modeling Language (SysML). SysML provides a framework and set of simulation tools for establishing and validating network support requirements for MMA mission concepts. Network infrastructure engineers affiliated with the MMA SSC will use SysML to facilitate development of new MMA science concepts from the community. \textit{Reproduced from \citet{roberts2021evaluation}.}}
    \label{fig:sysml}
\end{figure}

The space communications and navigation network infrastructure contributes to the overall feasibility and viability of an MMA mission concept. The signals, data, and reference frames provided by the network infrastructure are instrumental in establishing the time of observations, localization of the MMA source, and the target visibilities of the involved space-based observatories.  Additionally, MMA sources may emerge randomly and trigger unplanned demand for low-latency multi-point space data transport services and highly-automated spacecraft navigation, guidance and control responses.  Goddard’s network infrastructure engineers have established a Systems Modeling Language (SysML) framework to facilitate development of new MMA concepts and assist MMA community members in assessing the suitability of a network implementation for their science objectives. Figure~\ref{fig:sysml} provides an illustrative top-level structural decomposition of the elements involved in a transient science MMA mission concept using SysML.  Additional SysML diagrams describe and allocate functions to the MMA system elements. The network signals and data flows enabling the functions of the MMA system elements are defined. Subsequently, executable simulations are developed to determine and validate network support requirements for an MMA mission concept.

As a key contributor to the MMA SSC, the Goddard space communications and navigation community shall:
\begin{itemize}
    \item Provide communications and navigation content and links to informational resources for the MMA \ac{CAP}. 
    \item Assess the utility and feasibility of extending the improved MMA alert system data stream directly to space-based observatories using a government Tracking and Data Relay Satellite (TDRS) or commercial space broadcast service. 
    \item Provide tools for estimating network latency and other salient network suitability metrics to the MMA proposer tools catalogue. 
    \item Actively engage as part of the MMA community through the SSC technical forums, workshops, training events and seminars.
    \item Inform MMA science, engineering and operations stakeholders of relevant communications and navigation research, technologies, standards, and best practices.
    \item Elicit MMA community needs to align the communications and navigation capability roadmap. 
\end{itemize}

\section{A Typical Day at the MMA SSC}
\label{sec-typical_day}

During a future run, LIGO/Virgo issues an alert announcing the detection of \acp{GW} from a compact object merger. The alert and localization region for the merger event are received at the MMA SSC alert center. 

The MMA SSC receives the data from the GW alert and---using dedicated software and computational resources provided by the MMA SSC---cross correlates the localization region of the GW source with data from the Fermi GBM and Swift BAT, and possibly other observatories, in a search for coincident gamma-ray signals. Through a future version of the GCN, a community alert is issued announcing the detection of a gamma-ray EM counterpart to the GW source. The on duty staff then opens up the CAP Community Board for scientists and amateur astronomers to post information about their planned follow-up observations to search for EM emission at other wavelengths. Coordination of the various observatories on the ground and in space is facilitated by the MMA SSC scientists, who help observers run the Target Visibility Planner tool to determine the windows for observing with their instruments.   

As the EM follow-up planning starts, decisions need to be made in order to optimize and coordinate the use of multiple facilities and initiate Target of Opportunity requests for additional observations. The MMA SSC scientists serve as a reference point for facilitating these decisions among parties and communicating them to the community. The scientists on duty provide value-added information by providing specific expertise in the capabilities of NASA missions in the context of the EM follow-up of the GW event. This information is used by the community to optimize the follow-up strategy, with assistance from the MMA SSC scientists. 

After EM observations are initiated, the MMA SSC scientists work with the community by monitoring the observations and collecting data from all the NASA Missions at a central MMA SSC repository, so that interested members of the community can easily find them all in one place for ease of access and analysis.  Based on the outcome of the preliminary analysis, community members may request changes to the observation strategy. MMA SSC staff provide advice on the new course and work to revise the schedule and milestones, facilitate additional ToO requests, and communicate the changes to the broader community. Finally, the EM data are fully archived at the HEASARC for legacy community access. Using the HEASARC provided analysis tools, scientists around the world analyze the available observations and incorporate them in their work. 

Meanwhile, other MMA SSC scientists are processing the results of nightly observations by Rubin. Using the powerful AI/ML algorithms developed in house, they work to identify high-energy counterparts to the Rubin transients that may be MMA source candidates and produce a catalog to be disseminated to the community for analysis, which could include deep searches in GW and neutrino data to identify other messengers. The catalog contains sky coordinates, optical flux, possible source classifications, and archival information on high-energy and other emission  (e.g., host galaxy information from Roman data). 

During this time, the MMA GOF is assisting other users in developing proposals for EM follow-up of high-energy neutrino alerts from IceCube. The MMA GOF and MMA SSC services facilitate a new collaboration between high-energy astronomers interested in Swift X-ray follow-up of candidate blazar neutrino counterparts and radio astronomers with an active blazar monitoring program in the Southern hemisphere. 

Additionally, MMA SSC scientists are working with a proposal team developing a new mission concept for a next generation wide-field gamma-ray monitor. Based on training the MMA SSC sponsored several months prior, MMA SSC scientists are facilitating a student collaboration between the proposal team at a NASA Center and undergraduates at a minority-serving institution for the students to provide a ML based classification pipeline for on-board triggers.

Members of the MMA SSC staff are also working with mission PIs across the MMA community on science definition activities in support of future mission concepts and proposals. These activities include: (1) performing theoretical calculations and simulations to determine the science reach for a proposed mission concept; (2) assessing the physical implications of observations from the proposed mission concept within the context of observations from operating missions; and (3) assisting with science case development for proposals.

\section{Conclusions}
\label{sec-conclusions}

MMA is a burgeoning field but community coordination, collaboration, and communication (the 3Cs) are paramount to reap the most benefit and optimize resources. These 3Cs are at the heart of the MMA SSC. 
NASA is building a virtual MMA Science Support Center across GSFC and MSFC for 100\% community service. The MMA SSC will be the missing nexus between the ground and space communities, and among various space capabilities. Our Center leaders understand the importance of the SSC and have committed internal resources to seed some core functions.  More secure funding will be required to establish the SSC on a firm footing. 
The idea of a central hub at NASA for coordinating and facilitating MMA science has the support of several critical communities. We have received endorsement from LIGO/Virgo, IceCube, the Rubin Collaboration (Transient and Variable Star Collaboration, Science Coordination Group), National Radio Astronomy Observatory (NRAO), and several prominent leaders of MMA science. We look forward to collaborating with other national and international community-oriented services, as well as with academic institutions and industry, to better serve the MMA community and foster great science in the next decade and beyond.

\section*{Acknowledgements}
The MMA SSC Team acknowledges support from a Science Task Group award from the Science and Exploration Directorate at GSFC and support from the Science and Technology Office at MSFC.

\bibliographystyle{elsarticle-harv} 
\bibliography{main}

\begin{thebibliography}{29}
\expandafter\ifx\csname natexlab\endcsname\relax\def\natexlab#1{#1}\fi
\providecommand{\url}[1]{\texttt{#1}}
\providecommand{\href}[2]{#2}
\providecommand{\path}[1]{#1}
\providecommand{\DOIprefix}{doi:}
\providecommand{\ArXivprefix}{arXiv:}
\providecommand{\URLprefix}{URL: }
\providecommand{\Pubmedprefix}{pmid:}
\providecommand{\doi}[1]{\href{http://dx.doi.org/#1}{\path{#1}}}
\providecommand{\Pubmed}[1]{\href{pmid:#1}{\path{#1}}}
\providecommand{\bibinfo}[2]{#2}
\ifx\xfnm\relax \def\xfnm[#1]{\unskip,\space#1}\fi
\bibitem[{{Aartsen} et~al.(2021){Aartsen}, {Abbasi}, {Ackermann}, {Adams},
  {Aguilar}, {Ahlers}, {Ahrens}, {Alispach}, {Allison}, {Amin} and
  et~al.}]{2021JPhG...48f0501A}
\bibinfo{author}{{Aartsen}, M.G.}, \bibinfo{author}{{Abbasi}, R.},
  \bibinfo{author}{{Ackermann}, M.}, \bibinfo{author}{{Adams}, J.},
  \bibinfo{author}{{Aguilar}, J.A.}, \bibinfo{author}{{Ahlers}, M.},
  \bibinfo{author}{{Ahrens}, M.}, \bibinfo{author}{{Alispach}, C.},
  \bibinfo{author}{{Allison}, P.}, \bibinfo{author}{{Amin}, N.M.},
  \bibinfo{author}{et~al.}, \bibinfo{year}{2021}.
\newblock \bibinfo{title}{{IceCube-Gen2: the window to the extreme Universe}}.
\newblock \bibinfo{journal}{Journal of Physics G Nuclear Physics}
  \bibinfo{volume}{48}, \bibinfo{pages}{060501}.
\newblock \DOIprefix\doi{10.1088/1361-6471/abbd48}.
\bibitem[{{Aartsen} et~al.(2017){Aartsen}, {Ackermann}, {Adams}, {Aguilar},
  {Ahlers}, {Ahrens}, {Altmann}, {Andeen}, {Anderson}, {Ansseau} and
  et~al.}]{2017JInst..12P3012A}
\bibinfo{author}{{Aartsen}, M.G.}, \bibinfo{author}{{Ackermann}, M.},
  \bibinfo{author}{{Adams}, J.}, \bibinfo{author}{{Aguilar}, J.A.},
  \bibinfo{author}{{Ahlers}, M.}, \bibinfo{author}{{Ahrens}, M.},
  \bibinfo{author}{{Altmann}, D.}, \bibinfo{author}{{Andeen}, K.},
  \bibinfo{author}{{Anderson}, T.}, \bibinfo{author}{{Ansseau}, I.},
  \bibinfo{author}{et~al.}, \bibinfo{year}{2017}.
\newblock \bibinfo{title}{{The IceCube Neutrino Observatory: instrumentation
  and online systems}}.
\newblock \bibinfo{journal}{Journal of Instrumentation} \bibinfo{volume}{12},
  \bibinfo{pages}{P03012}.
\newblock \DOIprefix\doi{10.1088/1748-0221/12/03/P03012},
  \href{http://arxiv.org/abs/1612.05093}{{\tt arXiv:1612.05093}}.
\bibitem[{{Abbott} et~al.(2017){Abbott}, {Abbott}, {Abbott}, {Acernese},
  {Ackley}, {Adams}, {Adams}, {Addesso}, {Adhikari}, {Adya} and
  et~al.}]{2017ApJ...848L..12A}
\bibinfo{author}{{Abbott}, B.P.}, \bibinfo{author}{{Abbott}, R.},
  \bibinfo{author}{{Abbott}, T.D.}, \bibinfo{author}{{Acernese}, F.},
  \bibinfo{author}{{Ackley}, K.}, \bibinfo{author}{{Adams}, C.},
  \bibinfo{author}{{Adams}, T.}, \bibinfo{author}{{Addesso}, P.},
  \bibinfo{author}{{Adhikari}, R.X.}, \bibinfo{author}{{Adya}, V.B.},
  \bibinfo{author}{et~al.}, \bibinfo{year}{2017}.
\newblock \bibinfo{title}{{Multi-messenger Observations of a Binary Neutron
  Star Merger}}.
\newblock \bibinfo{journal}{\apjl} \bibinfo{volume}{848}, \bibinfo{pages}{L12}.
\newblock \DOIprefix\doi{10.3847/2041-8213/aa91c9},
  \href{http://arxiv.org/abs/1710.05833}{{\tt arXiv:1710.05833}}.
\bibitem[{{Acernese} et~al.(2015){Acernese}, {Agathos}, {Agatsuma}, {Aisa},
  {Allemandou}, {Allocca}, {Amarni}, {Astone}, {Balestri}, {Ballardin} and
  et~al.}]{2015CQGra..32b4001A}
\bibinfo{author}{{Acernese}, F.}, \bibinfo{author}{{Agathos}, M.},
  \bibinfo{author}{{Agatsuma}, K.}, \bibinfo{author}{{Aisa}, D.},
  \bibinfo{author}{{Allemandou}, N.}, \bibinfo{author}{{Allocca}, A.},
  \bibinfo{author}{{Amarni}, J.}, \bibinfo{author}{{Astone}, P.},
  \bibinfo{author}{{Balestri}, G.}, \bibinfo{author}{{Ballardin}, G.},
  \bibinfo{author}{et~al.}, \bibinfo{year}{2015}.
\newblock \bibinfo{title}{{Advanced Virgo: a second-generation interferometric
  gravitational wave detector}}.
\newblock \bibinfo{journal}{Classical and Quantum Gravity}
  \bibinfo{volume}{32}, \bibinfo{pages}{024001}.
\newblock \DOIprefix\doi{10.1088/0264-9381/32/2/024001},
  \href{http://arxiv.org/abs/1408.3978}{{\tt arXiv:1408.3978}}.
\bibitem[{{Ackermann} et~al.(2012){Ackermann}, {Ajello}, {Albert}, {Allafort},
  {Atwood}, {Axelsson}, {Baldini}, {Ballet}, {Barbiellini}, {Bastieri},
  {Bechtol}, {Bellazzini}, {Bissaldi}, {Blandford}, {Bloom}, {Bogart},
  {Bonamente}, {Borgland}, {Bottacini}, {Bouvier}, {Brandt}, {Bregeon},
  {Brigida}, {Bruel}, {Buehler}, {Burnett}, {Buson}, {Caliandro}, {Cameron},
  {Caraveo}, {Casandjian}, {Cavazzuti}, {Cecchi}, {{\c{C}}elik}, {Charles},
  {Chaves}, {Chekhtman}, {Cheung}, {Chiang}, {Ciprini}, {Claus},
  {Cohen-Tanugi}, {Conrad}, {Corbet}, {Cutini}, {D'Ammando}, {Davis}, {de
  Angelis}, {DeKlotz}, {de Palma}, {Dermer}, {Digel}, {Silva}, {Drell},
  {Drlica-Wagner}, {Dubois}, {Favuzzi}, {Fegan}, {Ferrara}, {Focke}, {Fortin},
  {Fukazawa}, {Funk}, {Fusco}, {Gargano}, {Gasparrini}, {Gehrels}, {Giebels},
  {Giglietto}, {Giordano}, {Giroletti}, {Glanzman}, {Godfrey}, {Grenier},
  {Grove}, {Guiriec}, {Hadasch}, {Hayashida}, {Hays}, {Horan}, {Hou}, {Hughes},
  {Jackson}, {Jogler}, {J{\'o}hannesson}, {Johnson}, {Johnson}, {Johnson},
  {Kamae}, {Katagiri}, {Kataoka}, {Kerr}, {Kn{\"o}dlseder}, {Kuss}, {Lande},
  {Larsson}, {Latronico}, {Lavalley}, {Lemoine-Goumard}, {Longo}, {Loparco},
  {Lott}, {Lovellette}, {Lubrano}, {Mazziotta}, {McConville}, {McEnery},
  {Mehault}, {Michelson}, {Mitthumsiri}, {Mizuno}, {Moiseev}, {Monte},
  {Monzani}, {Morselli}, {Moskalenko}, {Murgia}, {Naumann-Godo}, {Nemmen},
  {Nishino}, {Norris}, {Nuss}, {Ohno}, {Ohsugi}, {Okumura}, {Omodei},
  {Orienti}, {Orlando}, {Ormes}, {Paneque}, {Panetta}, {Perkins},
  {Pesce-Rollins}, {Pierbattista}, {Piron}, {Pivato}, {Porter}, {Racusin},
  {Rain{\`o}}, {Rando}, {Razzano}, {Razzaque}, {Reimer}, {Reimer}, {Reposeur},
  {Reyes}, {Ritz}, {Rochester}, {Romoli}, {Roth}, {Sadrozinski}, {Sanchez},
  {Saz Parkinson}, {Sbarra}, {Scargle}, {Sgr{\`o}}, {Siegal-Gaskins},
  {Siskind}, {Spandre}, {Spinelli}, {Stephens}, {Suson}, {Tajima}, {Takahashi},
  {Tanaka}, {Thayer}, {Thayer}, {Thompson}, {Tibaldo}, {Tinivella}, {Tosti},
  {Troja}, {Usher}, {Vandenbroucke}, {Van Klaveren}, {Vasileiou}, {Vianello},
  {Vitale}, {Waite}, {Wallace}, {Winer}, {Wood}, {Wood}, {Wood}, {Yang} and
  {Zimmer}}]{2012ApJS..203....4A}
\bibinfo{author}{{Ackermann}, M.}, \bibinfo{author}{{Ajello}, M.},
  \bibinfo{author}{{Albert}, A.}, \bibinfo{author}{{Allafort}, A.},
  \bibinfo{author}{{Atwood}, W.B.}, \bibinfo{author}{{Axelsson}, M.},
  \bibinfo{author}{{Baldini}, L.}, \bibinfo{author}{{Ballet}, J.},
  \bibinfo{author}{{Barbiellini}, G.}, \bibinfo{author}{{Bastieri}, D.},
  \bibinfo{author}{{Bechtol}, K.}, \bibinfo{author}{{Bellazzini}, R.},
  \bibinfo{author}{{Bissaldi}, E.}, \bibinfo{author}{{Blandford}, R.D.},
  \bibinfo{author}{{Bloom}, E.D.}, \bibinfo{author}{{Bogart}, J.R.},
  \bibinfo{author}{{Bonamente}, E.}, \bibinfo{author}{{Borgland}, A.W.},
  \bibinfo{author}{{Bottacini}, E.}, \bibinfo{author}{{Bouvier}, A.},
  \bibinfo{author}{{Brandt}, T.J.}, \bibinfo{author}{{Bregeon}, J.},
  \bibinfo{author}{{Brigida}, M.}, \bibinfo{author}{{Bruel}, P.},
  \bibinfo{author}{{Buehler}, R.}, \bibinfo{author}{{Burnett}, T.H.},
  \bibinfo{author}{{Buson}, S.}, \bibinfo{author}{{Caliandro}, G.A.},
  \bibinfo{author}{{Cameron}, R.A.}, \bibinfo{author}{{Caraveo}, P.A.},
  \bibinfo{author}{{Casandjian}, J.M.}, \bibinfo{author}{{Cavazzuti}, E.},
  \bibinfo{author}{{Cecchi}, C.}, \bibinfo{author}{{{\c{C}}elik}, {\"O}.},
  \bibinfo{author}{{Charles}, E.}, \bibinfo{author}{{Chaves}, R.C.G.},
  \bibinfo{author}{{Chekhtman}, A.}, \bibinfo{author}{{Cheung}, C.C.},
  \bibinfo{author}{{Chiang}, J.}, \bibinfo{author}{{Ciprini}, S.},
  \bibinfo{author}{{Claus}, R.}, \bibinfo{author}{{Cohen-Tanugi}, J.},
  \bibinfo{author}{{Conrad}, J.}, \bibinfo{author}{{Corbet}, R.},
  \bibinfo{author}{{Cutini}, S.}, \bibinfo{author}{{D'Ammando}, F.},
  \bibinfo{author}{{Davis}, D.S.}, \bibinfo{author}{{de Angelis}, A.},
  \bibinfo{author}{{DeKlotz}, M.}, \bibinfo{author}{{de Palma}, F.},
  \bibinfo{author}{{Dermer}, C.D.}, \bibinfo{author}{{Digel}, S.W.},
  \bibinfo{author}{{Silva}, E.d.C.e.}, \bibinfo{author}{{Drell}, P.S.},
  \bibinfo{author}{{Drlica-Wagner}, A.}, \bibinfo{author}{{Dubois}, R.},
  \bibinfo{author}{{Favuzzi}, C.}, \bibinfo{author}{{Fegan}, S.J.},
  \bibinfo{author}{{Ferrara}, E.C.}, \bibinfo{author}{{Focke}, W.B.},
  \bibinfo{author}{{Fortin}, P.}, \bibinfo{author}{{Fukazawa}, Y.},
  \bibinfo{author}{{Funk}, S.}, \bibinfo{author}{{Fusco}, P.},
  \bibinfo{author}{{Gargano}, F.}, \bibinfo{author}{{Gasparrini}, D.},
  \bibinfo{author}{{Gehrels}, N.}, \bibinfo{author}{{Giebels}, B.},
  \bibinfo{author}{{Giglietto}, N.}, \bibinfo{author}{{Giordano}, F.},
  \bibinfo{author}{{Giroletti}, M.}, \bibinfo{author}{{Glanzman}, T.},
  \bibinfo{author}{{Godfrey}, G.}, \bibinfo{author}{{Grenier}, I.A.},
  \bibinfo{author}{{Grove}, J.E.}, \bibinfo{author}{{Guiriec}, S.},
  \bibinfo{author}{{Hadasch}, D.}, \bibinfo{author}{{Hayashida}, M.},
  \bibinfo{author}{{Hays}, E.}, \bibinfo{author}{{Horan}, D.},
  \bibinfo{author}{{Hou}, X.}, \bibinfo{author}{{Hughes}, R.E.},
  \bibinfo{author}{{Jackson}, M.S.}, \bibinfo{author}{{Jogler}, T.},
  \bibinfo{author}{{J{\'o}hannesson}, G.}, \bibinfo{author}{{Johnson}, R.P.},
  \bibinfo{author}{{Johnson}, T.J.}, \bibinfo{author}{{Johnson}, W.N.},
  \bibinfo{author}{{Kamae}, T.}, \bibinfo{author}{{Katagiri}, H.},
  \bibinfo{author}{{Kataoka}, J.}, \bibinfo{author}{{Kerr}, M.},
  \bibinfo{author}{{Kn{\"o}dlseder}, J.}, \bibinfo{author}{{Kuss}, M.},
  \bibinfo{author}{{Lande}, J.}, \bibinfo{author}{{Larsson}, S.},
  \bibinfo{author}{{Latronico}, L.}, \bibinfo{author}{{Lavalley}, C.},
  \bibinfo{author}{{Lemoine-Goumard}, M.}, \bibinfo{author}{{Longo}, F.},
  \bibinfo{author}{{Loparco}, F.}, \bibinfo{author}{{Lott}, B.},
  \bibinfo{author}{{Lovellette}, M.N.}, \bibinfo{author}{{Lubrano}, P.},
  \bibinfo{author}{{Mazziotta}, M.N.}, \bibinfo{author}{{McConville}, W.},
  \bibinfo{author}{{McEnery}, J.E.}, \bibinfo{author}{{Mehault}, J.},
  \bibinfo{author}{{Michelson}, P.F.}, \bibinfo{author}{{Mitthumsiri}, W.},
  \bibinfo{author}{{Mizuno}, T.}, \bibinfo{author}{{Moiseev}, A.A.},
  \bibinfo{author}{{Monte}, C.}, \bibinfo{author}{{Monzani}, M.E.},
  \bibinfo{author}{{Morselli}, A.}, \bibinfo{author}{{Moskalenko}, I.V.},
  \bibinfo{author}{{Murgia}, S.}, \bibinfo{author}{{Naumann-Godo}, M.},
  \bibinfo{author}{{Nemmen}, R.}, \bibinfo{author}{{Nishino}, S.},
  \bibinfo{author}{{Norris}, J.P.}, \bibinfo{author}{{Nuss}, E.},
  \bibinfo{author}{{Ohno}, M.}, \bibinfo{author}{{Ohsugi}, T.},
  \bibinfo{author}{{Okumura}, A.}, \bibinfo{author}{{Omodei}, N.},
  \bibinfo{author}{{Orienti}, M.}, \bibinfo{author}{{Orlando}, E.},
  \bibinfo{author}{{Ormes}, J.F.}, \bibinfo{author}{{Paneque}, D.},
  \bibinfo{author}{{Panetta}, J.H.}, \bibinfo{author}{{Perkins}, J.S.},
  \bibinfo{author}{{Pesce-Rollins}, M.}, \bibinfo{author}{{Pierbattista}, M.},
  \bibinfo{author}{{Piron}, F.}, \bibinfo{author}{{Pivato}, G.},
  \bibinfo{author}{{Porter}, T.A.}, \bibinfo{author}{{Racusin}, J.L.},
  \bibinfo{author}{{Rain{\`o}}, S.}, \bibinfo{author}{{Rando}, R.},
  \bibinfo{author}{{Razzano}, M.}, \bibinfo{author}{{Razzaque}, S.},
  \bibinfo{author}{{Reimer}, A.}, \bibinfo{author}{{Reimer}, O.},
  \bibinfo{author}{{Reposeur}, T.}, \bibinfo{author}{{Reyes}, L.C.},
  \bibinfo{author}{{Ritz}, S.}, \bibinfo{author}{{Rochester}, L.S.},
  \bibinfo{author}{{Romoli}, C.}, \bibinfo{author}{{Roth}, M.},
  \bibinfo{author}{{Sadrozinski}, H.F.W.}, \bibinfo{author}{{Sanchez}, D.A.},
  \bibinfo{author}{{Saz Parkinson}, P.M.}, \bibinfo{author}{{Sbarra}, C.},
  \bibinfo{author}{{Scargle}, J.D.}, \bibinfo{author}{{Sgr{\`o}}, C.},
  \bibinfo{author}{{Siegal-Gaskins}, J.}, \bibinfo{author}{{Siskind}, E.J.},
  \bibinfo{author}{{Spandre}, G.}, \bibinfo{author}{{Spinelli}, P.},
  \bibinfo{author}{{Stephens}, T.E.}, \bibinfo{author}{{Suson}, D.J.},
  \bibinfo{author}{{Tajima}, H.}, \bibinfo{author}{{Takahashi}, H.},
  \bibinfo{author}{{Tanaka}, T.}, \bibinfo{author}{{Thayer}, J.G.},
  \bibinfo{author}{{Thayer}, J.B.}, \bibinfo{author}{{Thompson}, D.J.},
  \bibinfo{author}{{Tibaldo}, L.}, \bibinfo{author}{{Tinivella}, M.},
  \bibinfo{author}{{Tosti}, G.}, \bibinfo{author}{{Troja}, E.},
  \bibinfo{author}{{Usher}, T.L.}, \bibinfo{author}{{Vandenbroucke}, J.},
  \bibinfo{author}{{Van Klaveren}, B.}, \bibinfo{author}{{Vasileiou}, V.},
  \bibinfo{author}{{Vianello}, G.}, \bibinfo{author}{{Vitale}, V.},
  \bibinfo{author}{{Waite}, A.P.}, \bibinfo{author}{{Wallace}, E.},
  \bibinfo{author}{{Winer}, B.L.}, \bibinfo{author}{{Wood}, D.L.},
  \bibinfo{author}{{Wood}, K.S.}, \bibinfo{author}{{Wood}, M.},
  \bibinfo{author}{{Yang}, Z.}, \bibinfo{author}{{Zimmer}, S.},
  \bibinfo{year}{2012}.
\newblock \bibinfo{title}{{The Fermi Large Area Telescope on Orbit: Event
  Classification, Instrument Response Functions, and Calibration}}.
\newblock \bibinfo{journal}{\apjs} \bibinfo{volume}{203}, \bibinfo{pages}{4}.
\newblock \DOIprefix\doi{10.1088/0067-0049/203/1/4},
  \href{http://arxiv.org/abs/1206.1896}{{\tt arXiv:1206.1896}}.
\bibitem[{{Baker} et~al.(2019){Baker}, {Haiman}, {Rossi}, {Berger}, {Brandt},
  {Breedt}, {Breivik}, {Charisi}, {Derdzinski}, {D'Orazio}, {Ford}, {Greene},
  {Hill}, {Holley-Bockelmann}, {Key}, {Kocsis}, {Kupfer}, {Madau}, {Marsh},
  {McKernan}, {McWilliams}, {Natarajan}, {Nissanke}, {Noble}, {Phinney},
  {Ramsay}, {Schnittman}, {Sesana}, {Shoemaker}, {Stone}, {Toonen},
  {Trakhtenbrot}, {Vikhlinin} and {Volonteri}}]{2019BAAS...51c.123B}
\bibinfo{author}{{Baker}, J.}, \bibinfo{author}{{Haiman}, Z.},
  \bibinfo{author}{{Rossi}, E.M.}, \bibinfo{author}{{Berger}, E.},
  \bibinfo{author}{{Brandt}, N.}, \bibinfo{author}{{Breedt}, E.},
  \bibinfo{author}{{Breivik}, K.}, \bibinfo{author}{{Charisi}, M.},
  \bibinfo{author}{{Derdzinski}, A.}, \bibinfo{author}{{D'Orazio}, D.J.},
  \bibinfo{author}{{Ford}, S.}, \bibinfo{author}{{Greene}, J.E.},
  \bibinfo{author}{{Hill}, J.C.}, \bibinfo{author}{{Holley-Bockelmann}, K.},
  \bibinfo{author}{{Key}, J.S.}, \bibinfo{author}{{Kocsis}, B.},
  \bibinfo{author}{{Kupfer}, T.}, \bibinfo{author}{{Madau}, P.},
  \bibinfo{author}{{Marsh}, T.}, \bibinfo{author}{{McKernan}, B.},
  \bibinfo{author}{{McWilliams}, S.T.}, \bibinfo{author}{{Natarajan}, P.},
  \bibinfo{author}{{Nissanke}, S.}, \bibinfo{author}{{Noble}, S.},
  \bibinfo{author}{{Phinney}, E.S.}, \bibinfo{author}{{Ramsay}, G.},
  \bibinfo{author}{{Schnittman}, J.}, \bibinfo{author}{{Sesana}, A.},
  \bibinfo{author}{{Shoemaker}, D.}, \bibinfo{author}{{Stone}, N.},
  \bibinfo{author}{{Toonen}, S.}, \bibinfo{author}{{Trakhtenbrot}, B.},
  \bibinfo{author}{{Vikhlinin}, A.}, \bibinfo{author}{{Volonteri}, M.},
  \bibinfo{year}{2019}.
\newblock \bibinfo{title}{{Multimessenger science opportunities with mHz
  gravitational waves}}.
\newblock \bibinfo{journal}{\baas} \bibinfo{volume}{51}, \bibinfo{pages}{123}.
\newblock \href{http://arxiv.org/abs/1903.04417}{{\tt arXiv:1903.04417}}.
\bibitem[{{Bellm} et~al.(2019){Bellm}, {Kulkarni}, {Graham}, {Dekany}, {Smith},
  {Riddle}, {Masci}, {Helou}, {Prince}, {Adams}, {Barbarino}, {Barlow},
  {Bauer}, {Beck}, {Belicki}, {Biswas}, {Blagorodnova}, {Bodewits}, {Bolin},
  {Brinnel}, {Brooke}, {Bue}, {Bulla}, {Burruss}, {Cenko}, {Chang}, {Connolly},
  {Coughlin}, {Cromer}, {Cunningham}, {De}, {Delacroix}, {Desai}, {Duev},
  {Eadie}, {Farnham}, {Feeney}, {Feindt}, {Flynn}, {Franckowiak}, {Frederick},
  {Fremling}, {Gal-Yam}, {Gezari}, {Giomi}, {Goldstein}, {Golkhou}, {Goobar},
  {Groom}, {Hacopians}, {Hale}, {Henning}, {Ho}, {Hover}, {Howell}, {Hung},
  {Huppenkothen}, {Imel}, {Ip}, {Ivezi{\'c}}, {Jackson}, {Jones}, {Juric},
  {Kasliwal}, {Kaspi}, {Kaye}, {Kelley}, {Kowalski}, {Kramer}, {Kupfer},
  {Landry}, {Laher}, {Lee}, {Lin}, {Lin}, {Lunnan}, {Giomi}, {Mahabal}, {Mao},
  {Miller}, {Monkewitz}, {Murphy}, {Ngeow}, {Nordin}, {Nugent}, {Ofek},
  {Patterson}, {Penprase}, {Porter}, {Rauch}, {Rebbapragada}, {Reiley},
  {Rigault}, {Rodriguez}, {van Roestel}, {Rusholme}, {van Santen}, {Schulze},
  {Shupe}, {Singer}, {Soumagnac}, {Stein}, {Surace}, {Sollerman}, {Szkody},
  {Taddia}, {Terek}, {Van Sistine}, {van Velzen}, {Vestrand}, {Walters},
  {Ward}, {Ye}, {Yu}, {Yan} and {Zolkower}}]{2019PASP..131a8002B}
\bibinfo{author}{{Bellm}, E.C.}, \bibinfo{author}{{Kulkarni}, S.R.},
  \bibinfo{author}{{Graham}, M.J.}, \bibinfo{author}{{Dekany}, R.},
  \bibinfo{author}{{Smith}, R.M.}, \bibinfo{author}{{Riddle}, R.},
  \bibinfo{author}{{Masci}, F.J.}, \bibinfo{author}{{Helou}, G.},
  \bibinfo{author}{{Prince}, T.A.}, \bibinfo{author}{{Adams}, S.M.},
  \bibinfo{author}{{Barbarino}, C.}, \bibinfo{author}{{Barlow}, T.},
  \bibinfo{author}{{Bauer}, J.}, \bibinfo{author}{{Beck}, R.},
  \bibinfo{author}{{Belicki}, J.}, \bibinfo{author}{{Biswas}, R.},
  \bibinfo{author}{{Blagorodnova}, N.}, \bibinfo{author}{{Bodewits}, D.},
  \bibinfo{author}{{Bolin}, B.}, \bibinfo{author}{{Brinnel}, V.},
  \bibinfo{author}{{Brooke}, T.}, \bibinfo{author}{{Bue}, B.},
  \bibinfo{author}{{Bulla}, M.}, \bibinfo{author}{{Burruss}, R.},
  \bibinfo{author}{{Cenko}, S.B.}, \bibinfo{author}{{Chang}, C.K.},
  \bibinfo{author}{{Connolly}, A.}, \bibinfo{author}{{Coughlin}, M.},
  \bibinfo{author}{{Cromer}, J.}, \bibinfo{author}{{Cunningham}, V.},
  \bibinfo{author}{{De}, K.}, \bibinfo{author}{{Delacroix}, A.},
  \bibinfo{author}{{Desai}, V.}, \bibinfo{author}{{Duev}, D.A.},
  \bibinfo{author}{{Eadie}, G.}, \bibinfo{author}{{Farnham}, T.L.},
  \bibinfo{author}{{Feeney}, M.}, \bibinfo{author}{{Feindt}, U.},
  \bibinfo{author}{{Flynn}, D.}, \bibinfo{author}{{Franckowiak}, A.},
  \bibinfo{author}{{Frederick}, S.}, \bibinfo{author}{{Fremling}, C.},
  \bibinfo{author}{{Gal-Yam}, A.}, \bibinfo{author}{{Gezari}, S.},
  \bibinfo{author}{{Giomi}, M.}, \bibinfo{author}{{Goldstein}, D.A.},
  \bibinfo{author}{{Golkhou}, V.Z.}, \bibinfo{author}{{Goobar}, A.},
  \bibinfo{author}{{Groom}, S.}, \bibinfo{author}{{Hacopians}, E.},
  \bibinfo{author}{{Hale}, D.}, \bibinfo{author}{{Henning}, J.},
  \bibinfo{author}{{Ho}, A.Y.Q.}, \bibinfo{author}{{Hover}, D.},
  \bibinfo{author}{{Howell}, J.}, \bibinfo{author}{{Hung}, T.},
  \bibinfo{author}{{Huppenkothen}, D.}, \bibinfo{author}{{Imel}, D.},
  \bibinfo{author}{{Ip}, W.H.}, \bibinfo{author}{{Ivezi{\'c}}, {\v{Z}}.},
  \bibinfo{author}{{Jackson}, E.}, \bibinfo{author}{{Jones}, L.},
  \bibinfo{author}{{Juric}, M.}, \bibinfo{author}{{Kasliwal}, M.M.},
  \bibinfo{author}{{Kaspi}, S.}, \bibinfo{author}{{Kaye}, S.},
  \bibinfo{author}{{Kelley}, M.S.P.}, \bibinfo{author}{{Kowalski}, M.},
  \bibinfo{author}{{Kramer}, E.}, \bibinfo{author}{{Kupfer}, T.},
  \bibinfo{author}{{Landry}, W.}, \bibinfo{author}{{Laher}, R.R.},
  \bibinfo{author}{{Lee}, C.D.}, \bibinfo{author}{{Lin}, H.W.},
  \bibinfo{author}{{Lin}, Z.Y.}, \bibinfo{author}{{Lunnan}, R.},
  \bibinfo{author}{{Giomi}, M.}, \bibinfo{author}{{Mahabal}, A.},
  \bibinfo{author}{{Mao}, P.}, \bibinfo{author}{{Miller}, A.A.},
  \bibinfo{author}{{Monkewitz}, S.}, \bibinfo{author}{{Murphy}, P.},
  \bibinfo{author}{{Ngeow}, C.C.}, \bibinfo{author}{{Nordin}, J.},
  \bibinfo{author}{{Nugent}, P.}, \bibinfo{author}{{Ofek}, E.},
  \bibinfo{author}{{Patterson}, M.T.}, \bibinfo{author}{{Penprase}, B.},
  \bibinfo{author}{{Porter}, M.}, \bibinfo{author}{{Rauch}, L.},
  \bibinfo{author}{{Rebbapragada}, U.}, \bibinfo{author}{{Reiley}, D.},
  \bibinfo{author}{{Rigault}, M.}, \bibinfo{author}{{Rodriguez}, H.},
  \bibinfo{author}{{van Roestel}, J.}, \bibinfo{author}{{Rusholme}, B.},
  \bibinfo{author}{{van Santen}, J.}, \bibinfo{author}{{Schulze}, S.},
  \bibinfo{author}{{Shupe}, D.L.}, \bibinfo{author}{{Singer}, L.P.},
  \bibinfo{author}{{Soumagnac}, M.T.}, \bibinfo{author}{{Stein}, R.},
  \bibinfo{author}{{Surace}, J.}, \bibinfo{author}{{Sollerman}, J.},
  \bibinfo{author}{{Szkody}, P.}, \bibinfo{author}{{Taddia}, F.},
  \bibinfo{author}{{Terek}, S.}, \bibinfo{author}{{Van Sistine}, A.},
  \bibinfo{author}{{van Velzen}, S.}, \bibinfo{author}{{Vestrand}, W.T.},
  \bibinfo{author}{{Walters}, R.}, \bibinfo{author}{{Ward}, C.},
  \bibinfo{author}{{Ye}, Q.Z.}, \bibinfo{author}{{Yu}, P.C.},
  \bibinfo{author}{{Yan}, L.}, \bibinfo{author}{{Zolkower}, J.},
  \bibinfo{year}{2019}.
\newblock \bibinfo{title}{{The Zwicky Transient Facility: System Overview,
  Performance, and First Results}}.
\newblock \bibinfo{journal}{\pasp} \bibinfo{volume}{131},
  \bibinfo{pages}{018002}.
\newblock \DOIprefix\doi{10.1088/1538-3873/aaecbe},
  \href{http://arxiv.org/abs/1902.01932}{{\tt arXiv:1902.01932}}.
\bibitem[{{Burns}(2020)}]{2020LRR....23....4B}
\bibinfo{author}{{Burns}, E.}, \bibinfo{year}{2020}.
\newblock \bibinfo{title}{{Neutron star mergers and how to study them}}.
\newblock \bibinfo{journal}{Living Reviews in Relativity} \bibinfo{volume}{23},
  \bibinfo{pages}{4}.
\newblock \DOIprefix\doi{10.1007/s41114-020-00028-7},
  \href{http://arxiv.org/abs/1909.06085}{{\tt arXiv:1909.06085}}.
\bibitem[{{Dekany} et~al.(2020){Dekany}, {Smith}, {Riddle}, {Feeney}, {Porter},
  {Hale}, {Zolkower}, {Belicki}, {Kaye}, {Henning}, {Walters}, {Cromer},
  {Delacroix}, {Rodriguez}, {Reiley}, {Mao}, {Hover}, {Murphy}, {Burruss},
  {Baker}, {Kowalski}, {Reif}, {Mueller}, {Bellm}, {Graham} and
  {Kulkarni}}]{2020PASP..132c8001D}
\bibinfo{author}{{Dekany}, R.}, \bibinfo{author}{{Smith}, R.M.},
  \bibinfo{author}{{Riddle}, R.}, \bibinfo{author}{{Feeney}, M.},
  \bibinfo{author}{{Porter}, M.}, \bibinfo{author}{{Hale}, D.},
  \bibinfo{author}{{Zolkower}, J.}, \bibinfo{author}{{Belicki}, J.},
  \bibinfo{author}{{Kaye}, S.}, \bibinfo{author}{{Henning}, J.},
  \bibinfo{author}{{Walters}, R.}, \bibinfo{author}{{Cromer}, J.},
  \bibinfo{author}{{Delacroix}, A.}, \bibinfo{author}{{Rodriguez}, H.},
  \bibinfo{author}{{Reiley}, D.J.}, \bibinfo{author}{{Mao}, P.},
  \bibinfo{author}{{Hover}, D.}, \bibinfo{author}{{Murphy}, P.},
  \bibinfo{author}{{Burruss}, R.}, \bibinfo{author}{{Baker}, J.},
  \bibinfo{author}{{Kowalski}, M.}, \bibinfo{author}{{Reif}, K.},
  \bibinfo{author}{{Mueller}, P.}, \bibinfo{author}{{Bellm}, E.},
  \bibinfo{author}{{Graham}, M.}, \bibinfo{author}{{Kulkarni}, S.R.},
  \bibinfo{year}{2020}.
\newblock \bibinfo{title}{{The Zwicky Transient Facility: Observing System}}.
\newblock \bibinfo{journal}{\pasp} \bibinfo{volume}{132},
  \bibinfo{pages}{038001}.
\newblock \DOIprefix\doi{10.1088/1538-3873/ab4ca2},
  \href{http://arxiv.org/abs/2008.04923}{{\tt arXiv:2008.04923}}.
\bibitem[{{Gehrels} et~al.(2004){Gehrels}, {Chincarini}, {Giommi}, {Mason},
  {Nousek}, {Wells}, {White}, {Barthelmy}, {Burrows}, {Cominsky}, {Hurley},
  {Marshall}, {M{\'e}sz{\'a}ros}, {Roming}, {Angelini}, {Barbier}, {Belloni},
  {Campana}, {Caraveo}, {Chester}, {Citterio}, {Cline}, {Cropper}, {Cummings},
  {Dean}, {Feigelson}, {Fenimore}, {Frail}, {Fruchter}, {Garmire}, {Gendreau},
  {Ghisellini}, {Greiner}, {Hill}, {Hunsberger}, {Krimm}, {Kulkarni}, {Kumar},
  {Lebrun}, {Lloyd-Ronning}, {Markwardt}, {Mattson}, {Mushotzky}, {Norris},
  {Osborne}, {Paczynski}, {Palmer}, {Park}, {Parsons}, {Paul}, {Rees},
  {Reynolds}, {Rhoads}, {Sasseen}, {Schaefer}, {Short}, {Smale}, {Smith},
  {Stella}, {Tagliaferri}, {Takahashi}, {Tashiro}, {Townsley}, {Tueller},
  {Turner}, {Vietri}, {Voges}, {Ward}, {Willingale}, {Zerbi} and
  {Zhang}}]{2004ApJ...611.1005G}
\bibinfo{author}{{Gehrels}, N.}, \bibinfo{author}{{Chincarini}, G.},
  \bibinfo{author}{{Giommi}, P.}, \bibinfo{author}{{Mason}, K.O.},
  \bibinfo{author}{{Nousek}, J.A.}, \bibinfo{author}{{Wells}, A.A.},
  \bibinfo{author}{{White}, N.E.}, \bibinfo{author}{{Barthelmy}, S.D.},
  \bibinfo{author}{{Burrows}, D.N.}, \bibinfo{author}{{Cominsky}, L.R.},
  \bibinfo{author}{{Hurley}, K.C.}, \bibinfo{author}{{Marshall}, F.E.},
  \bibinfo{author}{{M{\'e}sz{\'a}ros}, P.}, \bibinfo{author}{{Roming}, P.W.A.},
  \bibinfo{author}{{Angelini}, L.}, \bibinfo{author}{{Barbier}, L.M.},
  \bibinfo{author}{{Belloni}, T.}, \bibinfo{author}{{Campana}, S.},
  \bibinfo{author}{{Caraveo}, P.A.}, \bibinfo{author}{{Chester}, M.M.},
  \bibinfo{author}{{Citterio}, O.}, \bibinfo{author}{{Cline}, T.L.},
  \bibinfo{author}{{Cropper}, M.S.}, \bibinfo{author}{{Cummings}, J.R.},
  \bibinfo{author}{{Dean}, A.J.}, \bibinfo{author}{{Feigelson}, E.D.},
  \bibinfo{author}{{Fenimore}, E.E.}, \bibinfo{author}{{Frail}, D.A.},
  \bibinfo{author}{{Fruchter}, A.S.}, \bibinfo{author}{{Garmire}, G.P.},
  \bibinfo{author}{{Gendreau}, K.}, \bibinfo{author}{{Ghisellini}, G.},
  \bibinfo{author}{{Greiner}, J.}, \bibinfo{author}{{Hill}, J.E.},
  \bibinfo{author}{{Hunsberger}, S.D.}, \bibinfo{author}{{Krimm}, H.A.},
  \bibinfo{author}{{Kulkarni}, S.R.}, \bibinfo{author}{{Kumar}, P.},
  \bibinfo{author}{{Lebrun}, F.}, \bibinfo{author}{{Lloyd-Ronning}, N.M.},
  \bibinfo{author}{{Markwardt}, C.B.}, \bibinfo{author}{{Mattson}, B.J.},
  \bibinfo{author}{{Mushotzky}, R.F.}, \bibinfo{author}{{Norris}, J.P.},
  \bibinfo{author}{{Osborne}, J.}, \bibinfo{author}{{Paczynski}, B.},
  \bibinfo{author}{{Palmer}, D.M.}, \bibinfo{author}{{Park}, H.S.},
  \bibinfo{author}{{Parsons}, A.M.}, \bibinfo{author}{{Paul}, J.},
  \bibinfo{author}{{Rees}, M.J.}, \bibinfo{author}{{Reynolds}, C.S.},
  \bibinfo{author}{{Rhoads}, J.E.}, \bibinfo{author}{{Sasseen}, T.P.},
  \bibinfo{author}{{Schaefer}, B.E.}, \bibinfo{author}{{Short}, A.T.},
  \bibinfo{author}{{Smale}, A.P.}, \bibinfo{author}{{Smith}, I.A.},
  \bibinfo{author}{{Stella}, L.}, \bibinfo{author}{{Tagliaferri}, G.},
  \bibinfo{author}{{Takahashi}, T.}, \bibinfo{author}{{Tashiro}, M.},
  \bibinfo{author}{{Townsley}, L.K.}, \bibinfo{author}{{Tueller}, J.},
  \bibinfo{author}{{Turner}, M.J.L.}, \bibinfo{author}{{Vietri}, M.},
  \bibinfo{author}{{Voges}, W.}, \bibinfo{author}{{Ward}, M.J.},
  \bibinfo{author}{{Willingale}, R.}, \bibinfo{author}{{Zerbi}, F.M.},
  \bibinfo{author}{{Zhang}, W.W.}, \bibinfo{year}{2004}.
\newblock \bibinfo{title}{{The Swift Gamma-Ray Burst Mission}}.
\newblock \bibinfo{journal}{\apj} \bibinfo{volume}{611},
  \bibinfo{pages}{1005--1020}.
\newblock \DOIprefix\doi{10.1086/422091},
  \href{http://arxiv.org/abs/astro-ph/0405233}{{\tt arXiv:astro-ph/0405233}}.
\bibitem[{{Gendreau} et~al.(2012){Gendreau}, {Arzoumanian} and
  {Okajima}}]{2012SPIE.8443E..13G}
\bibinfo{author}{{Gendreau}, K.C.}, \bibinfo{author}{{Arzoumanian}, Z.},
  \bibinfo{author}{{Okajima}, T.}, \bibinfo{year}{2012}.
\newblock \bibinfo{title}{{The Neutron star Interior Composition ExploreR
  (NICER): an Explorer mission of opportunity for soft x-ray timing
  spectroscopy}}, in: \bibinfo{editor}{{Takahashi}, T.},
  \bibinfo{editor}{{Murray}, S.S.}, \bibinfo{editor}{{den Herder}, J.W.A.}
  (Eds.), \bibinfo{booktitle}{Space Telescopes and Instrumentation 2012:
  Ultraviolet to Gamma Ray}, p. \bibinfo{pages}{844313}.
\newblock \DOIprefix\doi{10.1117/12.926396}.
\bibitem[{{Graham} et~al.(2019){Graham}, {Kulkarni}, {Bellm}, {Adams},
  {Barbarino}, {Blagorodnova}, {Bodewits}, {Bolin}, {Brady}, {Cenko}, {Chang},
  {Coughlin}, {De}, {Eadie}, {Farnham}, {Feindt}, {Franckowiak}, {Fremling},
  {Gezari}, {Ghosh}, {Goldstein}, {Golkhou}, {Goobar}, {Ho}, {Huppenkothen},
  {Ivezi{\'c}}, {Jones}, {Juric}, {Kaplan}, {Kasliwal}, {Kelley}, {Kupfer},
  {Lee}, {Lin}, {Lunnan}, {Mahabal}, {Miller}, {Ngeow}, {Nugent}, {Ofek},
  {Prince}, {Rauch}, {van Roestel}, {Schulze}, {Singer}, {Sollerman}, {Taddia},
  {Yan}, {Ye}, {Yu}, {Barlow}, {Bauer}, {Beck}, {Belicki}, {Biswas}, {Brinnel},
  {Brooke}, {Bue}, {Bulla}, {Burruss}, {Connolly}, {Cromer}, {Cunningham},
  {Dekany}, {Delacroix}, {Desai}, {Duev}, {Feeney}, {Flynn}, {Frederick},
  {Gal-Yam}, {Giomi}, {Groom}, {Hacopians}, {Hale}, {Helou}, {Henning},
  {Hover}, {Hillenbrand}, {Howell}, {Hung}, {Imel}, {Ip}, {Jackson}, {Kaspi},
  {Kaye}, {Kowalski}, {Kramer}, {Kuhn}, {Landry}, {Laher}, {Mao}, {Masci},
  {Monkewitz}, {Murphy}, {Nordin}, {Patterson}, {Penprase}, {Porter},
  {Rebbapragada}, {Reiley}, {Riddle}, {Rigault}, {Rodriguez}, {Rusholme}, {van
  Santen}, {Shupe}, {Smith}, {Soumagnac}, {Stein}, {Surace}, {Szkody}, {Terek},
  {Van Sistine}, {van Velzen}, {Vestrand}, {Walters}, {Ward}, {Zhang} and
  {Zolkower}}]{2019PASP..131g8001G}
\bibinfo{author}{{Graham}, M.J.}, \bibinfo{author}{{Kulkarni}, S.R.},
  \bibinfo{author}{{Bellm}, E.C.}, \bibinfo{author}{{Adams}, S.M.},
  \bibinfo{author}{{Barbarino}, C.}, \bibinfo{author}{{Blagorodnova}, N.},
  \bibinfo{author}{{Bodewits}, D.}, \bibinfo{author}{{Bolin}, B.},
  \bibinfo{author}{{Brady}, P.R.}, \bibinfo{author}{{Cenko}, S.B.},
  \bibinfo{author}{{Chang}, C.K.}, \bibinfo{author}{{Coughlin}, M.W.},
  \bibinfo{author}{{De}, K.}, \bibinfo{author}{{Eadie}, G.},
  \bibinfo{author}{{Farnham}, T.L.}, \bibinfo{author}{{Feindt}, U.},
  \bibinfo{author}{{Franckowiak}, A.}, \bibinfo{author}{{Fremling}, C.},
  \bibinfo{author}{{Gezari}, S.}, \bibinfo{author}{{Ghosh}, S.},
  \bibinfo{author}{{Goldstein}, D.A.}, \bibinfo{author}{{Golkhou}, V.Z.},
  \bibinfo{author}{{Goobar}, A.}, \bibinfo{author}{{Ho}, A.Y.Q.},
  \bibinfo{author}{{Huppenkothen}, D.}, \bibinfo{author}{{Ivezi{\'c}},
  {\v{Z}}.}, \bibinfo{author}{{Jones}, R.L.}, \bibinfo{author}{{Juric}, M.},
  \bibinfo{author}{{Kaplan}, D.L.}, \bibinfo{author}{{Kasliwal}, M.M.},
  \bibinfo{author}{{Kelley}, M.S.P.}, \bibinfo{author}{{Kupfer}, T.},
  \bibinfo{author}{{Lee}, C.D.}, \bibinfo{author}{{Lin}, H.W.},
  \bibinfo{author}{{Lunnan}, R.}, \bibinfo{author}{{Mahabal}, A.A.},
  \bibinfo{author}{{Miller}, A.A.}, \bibinfo{author}{{Ngeow}, C.C.},
  \bibinfo{author}{{Nugent}, P.}, \bibinfo{author}{{Ofek}, E.O.},
  \bibinfo{author}{{Prince}, T.A.}, \bibinfo{author}{{Rauch}, L.},
  \bibinfo{author}{{van Roestel}, J.}, \bibinfo{author}{{Schulze}, S.},
  \bibinfo{author}{{Singer}, L.P.}, \bibinfo{author}{{Sollerman}, J.},
  \bibinfo{author}{{Taddia}, F.}, \bibinfo{author}{{Yan}, L.},
  \bibinfo{author}{{Ye}, Q.Z.}, \bibinfo{author}{{Yu}, P.C.},
  \bibinfo{author}{{Barlow}, T.}, \bibinfo{author}{{Bauer}, J.},
  \bibinfo{author}{{Beck}, R.}, \bibinfo{author}{{Belicki}, J.},
  \bibinfo{author}{{Biswas}, R.}, \bibinfo{author}{{Brinnel}, V.},
  \bibinfo{author}{{Brooke}, T.}, \bibinfo{author}{{Bue}, B.},
  \bibinfo{author}{{Bulla}, M.}, \bibinfo{author}{{Burruss}, R.},
  \bibinfo{author}{{Connolly}, A.}, \bibinfo{author}{{Cromer}, J.},
  \bibinfo{author}{{Cunningham}, V.}, \bibinfo{author}{{Dekany}, R.},
  \bibinfo{author}{{Delacroix}, A.}, \bibinfo{author}{{Desai}, V.},
  \bibinfo{author}{{Duev}, D.A.}, \bibinfo{author}{{Feeney}, M.},
  \bibinfo{author}{{Flynn}, D.}, \bibinfo{author}{{Frederick}, S.},
  \bibinfo{author}{{Gal-Yam}, A.}, \bibinfo{author}{{Giomi}, M.},
  \bibinfo{author}{{Groom}, S.}, \bibinfo{author}{{Hacopians}, E.},
  \bibinfo{author}{{Hale}, D.}, \bibinfo{author}{{Helou}, G.},
  \bibinfo{author}{{Henning}, J.}, \bibinfo{author}{{Hover}, D.},
  \bibinfo{author}{{Hillenbrand}, L.A.}, \bibinfo{author}{{Howell}, J.},
  \bibinfo{author}{{Hung}, T.}, \bibinfo{author}{{Imel}, D.},
  \bibinfo{author}{{Ip}, W.H.}, \bibinfo{author}{{Jackson}, E.},
  \bibinfo{author}{{Kaspi}, S.}, \bibinfo{author}{{Kaye}, S.},
  \bibinfo{author}{{Kowalski}, M.}, \bibinfo{author}{{Kramer}, E.},
  \bibinfo{author}{{Kuhn}, M.}, \bibinfo{author}{{Landry}, W.},
  \bibinfo{author}{{Laher}, R.R.}, \bibinfo{author}{{Mao}, P.},
  \bibinfo{author}{{Masci}, F.J.}, \bibinfo{author}{{Monkewitz}, S.},
  \bibinfo{author}{{Murphy}, P.}, \bibinfo{author}{{Nordin}, J.},
  \bibinfo{author}{{Patterson}, M.T.}, \bibinfo{author}{{Penprase}, B.},
  \bibinfo{author}{{Porter}, M.}, \bibinfo{author}{{Rebbapragada}, U.},
  \bibinfo{author}{{Reiley}, D.}, \bibinfo{author}{{Riddle}, R.},
  \bibinfo{author}{{Rigault}, M.}, \bibinfo{author}{{Rodriguez}, H.},
  \bibinfo{author}{{Rusholme}, B.}, \bibinfo{author}{{van Santen}, J.},
  \bibinfo{author}{{Shupe}, D.L.}, \bibinfo{author}{{Smith}, R.M.},
  \bibinfo{author}{{Soumagnac}, M.T.}, \bibinfo{author}{{Stein}, R.},
  \bibinfo{author}{{Surace}, J.}, \bibinfo{author}{{Szkody}, P.},
  \bibinfo{author}{{Terek}, S.}, \bibinfo{author}{{Van Sistine}, A.},
  \bibinfo{author}{{van Velzen}, S.}, \bibinfo{author}{{Vestrand}, W.T.},
  \bibinfo{author}{{Walters}, R.}, \bibinfo{author}{{Ward}, C.},
  \bibinfo{author}{{Zhang}, C.}, \bibinfo{author}{{Zolkower}, J.},
  \bibinfo{year}{2019}.
\newblock \bibinfo{title}{{The Zwicky Transient Facility: Science Objectives}}.
\newblock \bibinfo{journal}{\pasp} \bibinfo{volume}{131},
  \bibinfo{pages}{078001}.
\newblock \DOIprefix\doi{10.1088/1538-3873/ab006c},
  \href{http://arxiv.org/abs/1902.01945}{{\tt arXiv:1902.01945}}.
\bibitem[{{IceCube Collaboration} et~al.(2018a){IceCube Collaboration},
  {Aartsen}, {Ackermann}, {Adams}, {Aguilar}, {Ahlers}, {Ahrens}, {Al Samarai},
  {Altmann}, {Andeen} and et~al.}]{2018Sci...361.1378I}
\bibinfo{author}{{IceCube Collaboration}}, \bibinfo{author}{{Aartsen}, M.G.},
  \bibinfo{author}{{Ackermann}, M.}, \bibinfo{author}{{Adams}, J.},
  \bibinfo{author}{{Aguilar}, J.A.}, \bibinfo{author}{{Ahlers}, M.},
  \bibinfo{author}{{Ahrens}, M.}, \bibinfo{author}{{Al Samarai}, I.},
  \bibinfo{author}{{Altmann}, D.}, \bibinfo{author}{{Andeen}, K.},
  \bibinfo{author}{et~al.}, \bibinfo{year}{2018}a.
\newblock \bibinfo{title}{{Multimessenger observations of a flaring blazar
  coincident with high-energy neutrino IceCube-170922A}}.
\newblock \bibinfo{journal}{Science} \bibinfo{volume}{361},
  \bibinfo{pages}{eaat1378}.
\newblock \DOIprefix\doi{10.1126/science.aat1378},
  \href{http://arxiv.org/abs/1807.08816}{{\tt arXiv:1807.08816}}.
\bibitem[{{IceCube Collaboration} et~al.(2018b){IceCube Collaboration},
  {Aartsen}, {Ackermann}, {Adams}, {Aguilar}, {Ahlers}, {Ahrens}, {Samarai},
  {Altmann}, {Andeen} and et~al.}]{2018Sci...361..147I}
\bibinfo{author}{{IceCube Collaboration}}, \bibinfo{author}{{Aartsen}, M.G.},
  \bibinfo{author}{{Ackermann}, M.}, \bibinfo{author}{{Adams}, J.},
  \bibinfo{author}{{Aguilar}, J.A.}, \bibinfo{author}{{Ahlers}, M.},
  \bibinfo{author}{{Ahrens}, M.}, \bibinfo{author}{{Samarai}, I.A.},
  \bibinfo{author}{{Altmann}, D.}, \bibinfo{author}{{Andeen}, K.},
  \bibinfo{author}{et~al.}, \bibinfo{year}{2018}b.
\newblock \bibinfo{title}{{Neutrino emission from the direction of the blazar
  TXS 0506+056 prior to the IceCube-170922A alert}}.
\newblock \bibinfo{journal}{Science} \bibinfo{volume}{361},
  \bibinfo{pages}{147--151}.
\newblock \DOIprefix\doi{10.1126/science.aat2890},
  \href{http://arxiv.org/abs/1807.08794}{{\tt arXiv:1807.08794}}.
\bibitem[{{Ivezi{\'c}} et~al.(2019){Ivezi{\'c}}, {Kahn}, {Tyson}, {Abel},
  {Acosta}, {Allsman}, {Alonso}, {AlSayyad}, {Anderson}, {Andrew} and
  et~al.}]{2019ApJ...873..111I}
\bibinfo{author}{{Ivezi{\'c}}, {\v{Z}}.}, \bibinfo{author}{{Kahn}, S.M.},
  \bibinfo{author}{{Tyson}, J.A.}, \bibinfo{author}{{Abel}, B.},
  \bibinfo{author}{{Acosta}, E.}, \bibinfo{author}{{Allsman}, R.},
  \bibinfo{author}{{Alonso}, D.}, \bibinfo{author}{{AlSayyad}, Y.},
  \bibinfo{author}{{Anderson}, S.F.}, \bibinfo{author}{{Andrew}, J.},
  \bibinfo{author}{et~al.}, \bibinfo{year}{2019}.
\newblock \bibinfo{title}{{LSST: From Science Drivers to Reference Design and
  Anticipated Data Products}}.
\newblock \bibinfo{journal}{\apj} \bibinfo{volume}{873}, \bibinfo{pages}{111}.
\newblock \DOIprefix\doi{10.3847/1538-4357/ab042c},
  \href{http://arxiv.org/abs/0805.2366}{{\tt arXiv:0805.2366}}.
\bibitem[{{Kuulkers} et~al.(2019){Kuulkers}, {Ehle}, {Gabriel}, {Ibarra},
  {Kretschmar}, {Mer{\'\i}n}, {Ness}, {Salazar}, {Salgado},
  {S{\'a}nchez-Fern{\'a}ndez}, {Saxton} and {Levesque}}]{2019ASPC..523..503K}
\bibinfo{author}{{Kuulkers}, E.}, \bibinfo{author}{{Ehle}, M.},
  \bibinfo{author}{{Gabriel}, C.}, \bibinfo{author}{{Ibarra}, A.},
  \bibinfo{author}{{Kretschmar}, P.}, \bibinfo{author}{{Mer{\'\i}n}, B.},
  \bibinfo{author}{{Ness}, J.U.}, \bibinfo{author}{{Salazar}, E.},
  \bibinfo{author}{{Salgado}, J.},
  \bibinfo{author}{{S{\'a}nchez-Fern{\'a}ndez}, C.}, \bibinfo{author}{{Saxton},
  R.}, \bibinfo{author}{{Levesque}, E.M.}, \bibinfo{year}{2019}.
\newblock \bibinfo{title}{{Coordinating Observations Among Ground and
  Space-Based Telescopes in the Multi-Messenger Era}}, in:
  \bibinfo{editor}{{Teuben}, P.J.}, \bibinfo{editor}{{Pound}, M.W.},
  \bibinfo{editor}{{Thomas}, B.A.}, \bibinfo{editor}{{Warner}, E.M.} (Eds.),
  \bibinfo{booktitle}{Astronomical Data Analysis Software and Systems XXVII},
  p. \bibinfo{pages}{503}.
\newblock \href{http://arxiv.org/abs/1901.05390}{{\tt arXiv:1901.05390}}.
\bibitem[{{LIGO Scientific Collaboration} et~al.(2015){LIGO Scientific
  Collaboration}, {Aasi}, {Abbott}, {Abbott}, {Abbott}, {Abernathy}, {Ackley},
  {Adams}, {Adams}, {Addesso} and et~al.}]{2015CQGra..32g4001L}
\bibinfo{author}{{LIGO Scientific Collaboration}}, \bibinfo{author}{{Aasi},
  J.}, \bibinfo{author}{{Abbott}, B.P.}, \bibinfo{author}{{Abbott}, R.},
  \bibinfo{author}{{Abbott}, T.}, \bibinfo{author}{{Abernathy}, M.R.},
  \bibinfo{author}{{Ackley}, K.}, \bibinfo{author}{{Adams}, C.},
  \bibinfo{author}{{Adams}, T.}, \bibinfo{author}{{Addesso}, P.},
  \bibinfo{author}{et~al.}, \bibinfo{year}{2015}.
\newblock \bibinfo{title}{{Advanced LIGO}}.
\newblock \bibinfo{journal}{Classical and Quantum Gravity}
  \bibinfo{volume}{32}, \bibinfo{pages}{074001}.
\newblock \DOIprefix\doi{10.1088/0264-9381/32/7/074001},
  \href{http://arxiv.org/abs/1411.4547}{{\tt arXiv:1411.4547}}.
\bibitem[{{Masci} et~al.(2019){Masci}, {Laher}, {Rusholme}, {Shupe}, {Groom},
  {Surace}, {Jackson}, {Monkewitz}, {Beck}, {Flynn}, {Terek}, {Landry},
  {Hacopians}, {Desai}, {Howell}, {Brooke}, {Imel}, {Wachter}, {Ye}, {Lin},
  {Cenko}, {Cunningham}, {Rebbapragada}, {Bue}, {Miller}, {Mahabal}, {Bellm},
  {Patterson}, {Juri{\'c}}, {Golkhou}, {Ofek}, {Walters}, {Graham}, {Kasliwal},
  {Dekany}, {Kupfer}, {Burdge}, {Cannella}, {Barlow}, {Van Sistine}, {Giomi},
  {Fremling}, {Blagorodnova}, {Levitan}, {Riddle}, {Smith}, {Helou}, {Prince}
  and {Kulkarni}}]{2019PASP..131a8003M}
\bibinfo{author}{{Masci}, F.J.}, \bibinfo{author}{{Laher}, R.R.},
  \bibinfo{author}{{Rusholme}, B.}, \bibinfo{author}{{Shupe}, D.L.},
  \bibinfo{author}{{Groom}, S.}, \bibinfo{author}{{Surace}, J.},
  \bibinfo{author}{{Jackson}, E.}, \bibinfo{author}{{Monkewitz}, S.},
  \bibinfo{author}{{Beck}, R.}, \bibinfo{author}{{Flynn}, D.},
  \bibinfo{author}{{Terek}, S.}, \bibinfo{author}{{Landry}, W.},
  \bibinfo{author}{{Hacopians}, E.}, \bibinfo{author}{{Desai}, V.},
  \bibinfo{author}{{Howell}, J.}, \bibinfo{author}{{Brooke}, T.},
  \bibinfo{author}{{Imel}, D.}, \bibinfo{author}{{Wachter}, S.},
  \bibinfo{author}{{Ye}, Q.Z.}, \bibinfo{author}{{Lin}, H.W.},
  \bibinfo{author}{{Cenko}, S.B.}, \bibinfo{author}{{Cunningham}, V.},
  \bibinfo{author}{{Rebbapragada}, U.}, \bibinfo{author}{{Bue}, B.},
  \bibinfo{author}{{Miller}, A.A.}, \bibinfo{author}{{Mahabal}, A.},
  \bibinfo{author}{{Bellm}, E.C.}, \bibinfo{author}{{Patterson}, M.T.},
  \bibinfo{author}{{Juri{\'c}}, M.}, \bibinfo{author}{{Golkhou}, V.Z.},
  \bibinfo{author}{{Ofek}, E.O.}, \bibinfo{author}{{Walters}, R.},
  \bibinfo{author}{{Graham}, M.}, \bibinfo{author}{{Kasliwal}, M.M.},
  \bibinfo{author}{{Dekany}, R.G.}, \bibinfo{author}{{Kupfer}, T.},
  \bibinfo{author}{{Burdge}, K.}, \bibinfo{author}{{Cannella}, C.B.},
  \bibinfo{author}{{Barlow}, T.}, \bibinfo{author}{{Van Sistine}, A.},
  \bibinfo{author}{{Giomi}, M.}, \bibinfo{author}{{Fremling}, C.},
  \bibinfo{author}{{Blagorodnova}, N.}, \bibinfo{author}{{Levitan}, D.},
  \bibinfo{author}{{Riddle}, R.}, \bibinfo{author}{{Smith}, R.M.},
  \bibinfo{author}{{Helou}, G.}, \bibinfo{author}{{Prince}, T.A.},
  \bibinfo{author}{{Kulkarni}, S.R.}, \bibinfo{year}{2019}.
\newblock \bibinfo{title}{{The Zwicky Transient Facility: Data Processing,
  Products, and Archive}}.
\newblock \bibinfo{journal}{\pasp} \bibinfo{volume}{131},
  \bibinfo{pages}{018003}.
\newblock \DOIprefix\doi{10.1088/1538-3873/aae8ac},
  \href{http://arxiv.org/abs/1902.01872}{{\tt arXiv:1902.01872}}.
\bibitem[{{Matheson} et~al.(2019){Matheson}, {Stubens}, {Soraisam}, {Narayan},
  {Saha}, {Lee}, {Wolf}, {Merrill}, {Ridgway}, {Bolton}, {Snodgrass},
  {Scheidegger}, {Kececioglu}, {Peek}, {Rest}, {Smith}, {Momcheva}, {Petravick}
  and {Morganson}}]{2019BAAS...51g.139M}
\bibinfo{author}{{Matheson}, T.}, \bibinfo{author}{{Stubens}, C.},
  \bibinfo{author}{{Soraisam}, M.}, \bibinfo{author}{{Narayan}, G.},
  \bibinfo{author}{{Saha}, A.}, \bibinfo{author}{{Lee}, C.H.},
  \bibinfo{author}{{Wolf}, N.}, \bibinfo{author}{{Merrill}, C.},
  \bibinfo{author}{{Ridgway}, S.}, \bibinfo{author}{{Bolton}, A.},
  \bibinfo{author}{{Snodgrass}, R.}, \bibinfo{author}{{Scheidegger}, C.},
  \bibinfo{author}{{Kececioglu}, J.}, \bibinfo{author}{{Peek}, J.},
  \bibinfo{author}{{Rest}, A.}, \bibinfo{author}{{Smith}, A.},
  \bibinfo{author}{{Momcheva}, I.}, \bibinfo{author}{{Petravick}, D.},
  \bibinfo{author}{{Morganson}, E.}, \bibinfo{year}{2019}.
\newblock \bibinfo{title}{{ANTARES: Enabling Time-Domain Discovery in the
  2020s}}, in: \bibinfo{booktitle}{Bulletin of the American Astronomical
  Society}, p. \bibinfo{pages}{139}.
\bibitem[{{McEnery} et~al.(2019){McEnery}, {van der Horst}, {Dominguez},
  {Moiseev}, {Marcowith}, {Harding}, {Lien}, {Giuliani}, {Inglis}, {Ansoldi},
  {Stamerra}, {Manousakis}, {Strong}, {Bambi}, {Patricelli}, {Baring},
  {Barrio}, {Bastieri}, {Fields}, {Beacom}, {Beckmann}, {Bednarek}, {Rani},
  {Boggs}, {Bolotnikov}, {Cenko}, {Buckley}, {Grefenstette}, {Hui}, {Pittori},
  {Prescod-Weinstein}, {Shrader}, {Gouiffes}, {Kierans}, {Wilson-Hodge},
  {D'Ammando}, {Castro}, {Kocveski}, {Gasparrini}, {Thompson}, {Williams}, {De
  Angelis}, {Bernard}, {Digel}, {Morcuende}, {Charles}, {Bissaldi}, {Hays},
  {Ferrara}, {Bozzo}, {Grove}, {Wulf}, {Bottacini}, {Caroli}, {Kislat},
  {Oikonomou}, {Giordano}, {Longo}, {Fryer}, {Fukazawa}, {Georganopoulos}, {De
  Nolfo}, {Vianello}, {Kanbach}, {Younes}, {Blumer}, {Hartmann}, {Hernanz},
  {Takahashi}, {Li}, {Agudo}, {Moskalenko}, {Stumke}, {Grenier}, {Smith},
  {Rodi}, {Perkins}, {Gelfand}, {Holder}, {Knodlseder}, {Kopp}, {Lenain},
  {{\'A}lvarez}, {Metcalfe}, {Krizmanic}, {Stephen}, {Hewitt}, {Mitchell},
  {Harding}, {Tomsick}, {Racusin}, {Finke}, {Kargaltsev}, {Klimenko},
  {Krawczynski}, {Smith}, {Kubo}, {Di Venere}, {Marcotulli}, {Lommler},
  {Parker}, {Baldini}, {Foffano}, {Zampieri}, {Tibaldo}, {Petropoulou},
  {Ajello}, {Meyer}, {L{\'o}pez}, {McConnell}, {Boettcher}, {Cardillo},
  {Martinez}, {Kerr}, {Mazziotta}, {McEnery}, {Di Mauro}, {Wood}, {Meyer},
  {Briggs}, {De Becker}, {Lovellette}, {Doro}, {Sanchez-Conde}, {Moss},
  {Mizuno}, {Rib{\'o}}, {Nakazawa}, {Neilson}, {Auricchio}, {Omodei},
  {Oberlack}, {Ohno}, {Orlando}, {Otte}, {Coppi}, {Bloser}, {Zhang}, {Laurent},
  {Pohl}, {Prandini}, {Shawhan}, {Caputo}, {Campana}, {Rando}, {Woolf},
  {Johnson}, {Mignani}, {Walter}, {Ojha}, {da Silva}, {Dietrich}, {Funk},
  {Zane}, {Anton}, {Buson}, {Cutini}, {Saz Parkinson}, {Schirato}, {Griffin},
  {Kaufmann}, {Stawarz}, {Ciprini}, {Del Sordo}, {Jones}, {Guiriec}, {Tajima},
  {Cheung}, {The}, {Venters}, {Porter}, {Linden}, {Barres}, {Paliya},
  {Bozhilov}, {Vestrand}, {Tatischeff}, {Chen}, {Wang}, {Tanaka}, {Uhm},
  {Zhang}, {Zimmer}, {Zoglauer} and {Wadiasingh}}]{2019BAAS...51g.245M}
\bibinfo{author}{{McEnery}, J.}, \bibinfo{author}{{van der Horst}, A.},
  \bibinfo{author}{{Dominguez}, A.}, \bibinfo{author}{{Moiseev}, A.},
  \bibinfo{author}{{Marcowith}, A.}, \bibinfo{author}{{Harding}, A.},
  \bibinfo{author}{{Lien}, A.}, \bibinfo{author}{{Giuliani}, A.},
  \bibinfo{author}{{Inglis}, A.}, \bibinfo{author}{{Ansoldi}, S.},
  \bibinfo{author}{{Stamerra}, A.}, \bibinfo{author}{{Manousakis}, A.},
  \bibinfo{author}{{Strong}, A.}, \bibinfo{author}{{Bambi}, C.},
  \bibinfo{author}{{Patricelli}, B.}, \bibinfo{author}{{Baring}, M.},
  \bibinfo{author}{{Barrio}, J.A.}, \bibinfo{author}{{Bastieri}, D.},
  \bibinfo{author}{{Fields}, B.}, \bibinfo{author}{{Beacom}, J.},
  \bibinfo{author}{{Beckmann}, V.}, \bibinfo{author}{{Bednarek}, W.},
  \bibinfo{author}{{Rani}, B.}, \bibinfo{author}{{Boggs}, S.},
  \bibinfo{author}{{Bolotnikov}, A.}, \bibinfo{author}{{Cenko}, S.B.},
  \bibinfo{author}{{Buckley}, J.}, \bibinfo{author}{{Grefenstette}, B.},
  \bibinfo{author}{{Hui}, M.}, \bibinfo{author}{{Pittori}, C.},
  \bibinfo{author}{{Prescod-Weinstein}, C.}, \bibinfo{author}{{Shrader}, C.},
  \bibinfo{author}{{Gouiffes}, C.}, \bibinfo{author}{{Kierans}, C.},
  \bibinfo{author}{{Wilson-Hodge}, C.}, \bibinfo{author}{{D'Ammando}, F.},
  \bibinfo{author}{{Castro}, D.}, \bibinfo{author}{{Kocveski}, D.},
  \bibinfo{author}{{Gasparrini}, D.}, \bibinfo{author}{{Thompson}, D.},
  \bibinfo{author}{{Williams}, D.}, \bibinfo{author}{{De Angelis}, A.},
  \bibinfo{author}{{Bernard}, D.}, \bibinfo{author}{{Digel}, S.},
  \bibinfo{author}{{Morcuende}, D.}, \bibinfo{author}{{Charles}, E.},
  \bibinfo{author}{{Bissaldi}, E.}, \bibinfo{author}{{Hays}, E.},
  \bibinfo{author}{{Ferrara}, E.}, \bibinfo{author}{{Bozzo}, E.},
  \bibinfo{author}{{Grove}, E.}, \bibinfo{author}{{Wulf}, E.},
  \bibinfo{author}{{Bottacini}, E.}, \bibinfo{author}{{Caroli}, E.},
  \bibinfo{author}{{Kislat}, F.}, \bibinfo{author}{{Oikonomou}, F.},
  \bibinfo{author}{{Giordano}, F.}, \bibinfo{author}{{Longo}, F.},
  \bibinfo{author}{{Fryer}, C.}, \bibinfo{author}{{Fukazawa}, Y.},
  \bibinfo{author}{{Georganopoulos}, M.}, \bibinfo{author}{{De Nolfo}, G.},
  \bibinfo{author}{{Vianello}, G.}, \bibinfo{author}{{Kanbach}, G.},
  \bibinfo{author}{{Younes}, G.}, \bibinfo{author}{{Blumer}, H.},
  \bibinfo{author}{{Hartmann}, D.}, \bibinfo{author}{{Hernanz}, M.},
  \bibinfo{author}{{Takahashi}, H.}, \bibinfo{author}{{Li}, H.},
  \bibinfo{author}{{Agudo}, I.}, \bibinfo{author}{{Moskalenko}, I.},
  \bibinfo{author}{{Stumke}, I.}, \bibinfo{author}{{Grenier}, I.},
  \bibinfo{author}{{Smith}, J.}, \bibinfo{author}{{Rodi}, J.},
  \bibinfo{author}{{Perkins}, J.}, \bibinfo{author}{{Gelfand}, J.},
  \bibinfo{author}{{Holder}, J.}, \bibinfo{author}{{Knodlseder}, J.},
  \bibinfo{author}{{Kopp}, J.}, \bibinfo{author}{{Lenain}, J.P.},
  \bibinfo{author}{{{\'A}lvarez}, J.M.}, \bibinfo{author}{{Metcalfe}, J.},
  \bibinfo{author}{{Krizmanic}, J.}, \bibinfo{author}{{Stephen}, J.B.},
  \bibinfo{author}{{Hewitt}, J.}, \bibinfo{author}{{Mitchell}, J.},
  \bibinfo{author}{{Harding}, P.}, \bibinfo{author}{{Tomsick}, J.},
  \bibinfo{author}{{Racusin}, J.}, \bibinfo{author}{{Finke}, J.},
  \bibinfo{author}{{Kargaltsev}, O.}, \bibinfo{author}{{Klimenko}, A.V.},
  \bibinfo{author}{{Krawczynski}, H.}, \bibinfo{author}{{Smith}, K.},
  \bibinfo{author}{{Kubo}, H.}, \bibinfo{author}{{Di Venere}, L.},
  \bibinfo{author}{{Marcotulli}, L.}, \bibinfo{author}{{Lommler}, J.},
  \bibinfo{author}{{Parker}, L.}, \bibinfo{author}{{Baldini}, L.},
  \bibinfo{author}{{Foffano}, L.}, \bibinfo{author}{{Zampieri}, L.},
  \bibinfo{author}{{Tibaldo}, L.}, \bibinfo{author}{{Petropoulou}, M.},
  \bibinfo{author}{{Ajello}, M.}, \bibinfo{author}{{Meyer}, M.},
  \bibinfo{author}{{L{\'o}pez}, M.}, \bibinfo{author}{{McConnell}, M.},
  \bibinfo{author}{{Boettcher}, M.}, \bibinfo{author}{{Cardillo}, M.},
  \bibinfo{author}{{Martinez}, M.}, \bibinfo{author}{{Kerr}, M.},
  \bibinfo{author}{{Mazziotta}, M.N.}, \bibinfo{author}{{McEnery}, J.},
  \bibinfo{author}{{Di Mauro}, M.}, \bibinfo{author}{{Wood}, M.},
  \bibinfo{author}{{Meyer}, E.}, \bibinfo{author}{{Briggs}, M.},
  \bibinfo{author}{{De Becker}, M.}, \bibinfo{author}{{Lovellette}, M.},
  \bibinfo{author}{{Doro}, M.}, \bibinfo{author}{{Sanchez-Conde}, M.A.},
  \bibinfo{author}{{Moss}, M.}, \bibinfo{author}{{Mizuno}, T.},
  \bibinfo{author}{{Rib{\'o}}, M.}, \bibinfo{author}{{Nakazawa}, K.},
  \bibinfo{author}{{Neilson}, N.K.}, \bibinfo{author}{{Auricchio}, N.},
  \bibinfo{author}{{Omodei}, N.}, \bibinfo{author}{{Oberlack}, U.},
  \bibinfo{author}{{Ohno}, M.}, \bibinfo{author}{{Orlando}, E.},
  \bibinfo{author}{{Otte}, N.}, \bibinfo{author}{{Coppi}, P.},
  \bibinfo{author}{{Bloser}, P.}, \bibinfo{author}{{Zhang}, H.},
  \bibinfo{author}{{Laurent}, P.}, \bibinfo{author}{{Pohl}, M.},
  \bibinfo{author}{{Prandini}, E.}, \bibinfo{author}{{Shawhan}, P.},
  \bibinfo{author}{{Caputo}, R.}, \bibinfo{author}{{Campana}, R.},
  \bibinfo{author}{{Rando}, R.}, \bibinfo{author}{{Woolf}, R.},
  \bibinfo{author}{{Johnson}, R.}, \bibinfo{author}{{Mignani}, R.},
  \bibinfo{author}{{Walter}, R.}, \bibinfo{author}{{Ojha}, R.},
  \bibinfo{author}{{da Silva}, R.C.}, \bibinfo{author}{{Dietrich}, S.},
  \bibinfo{author}{{Funk}, S.}, \bibinfo{author}{{Zane}, S.},
  \bibinfo{author}{{Anton}, S.}, \bibinfo{author}{{Buson}, S.},
  \bibinfo{author}{{Cutini}, S.}, \bibinfo{author}{{Saz Parkinson}, P.},
  \bibinfo{author}{{Schirato}, R.}, \bibinfo{author}{{Griffin}, S.},
  \bibinfo{author}{{Kaufmann}, S.}, \bibinfo{author}{{Stawarz}, L.},
  \bibinfo{author}{{Ciprini}, S.}, \bibinfo{author}{{Del Sordo}, S.},
  \bibinfo{author}{{Jones}, S.}, \bibinfo{author}{{Guiriec}, S.},
  \bibinfo{author}{{Tajima}, H.}, \bibinfo{author}{{Cheung}, T.},
  \bibinfo{author}{{The}, L.S.}, \bibinfo{author}{{Venters}, T.},
  \bibinfo{author}{{Porter}, T.}, \bibinfo{author}{{Linden}, T.},
  \bibinfo{author}{{Barres}, U.}, \bibinfo{author}{{Paliya}, V.S.},
  \bibinfo{author}{{Bozhilov}, V.}, \bibinfo{author}{{Vestrand}, T.},
  \bibinfo{author}{{Tatischeff}, V.}, \bibinfo{author}{{Chen}, W.},
  \bibinfo{author}{{Wang}, X.}, \bibinfo{author}{{Tanaka}, Y.},
  \bibinfo{author}{{Uhm}, L.}, \bibinfo{author}{{Zhang}, B.},
  \bibinfo{author}{{Zimmer}, S.}, \bibinfo{author}{{Zoglauer}, A.},
  \bibinfo{author}{{Wadiasingh}, Z.}, \bibinfo{year}{2019}.
\newblock \bibinfo{title}{{All-sky Medium Energy Gamma-ray Observatory:
  Exploring the Extreme Multimessenger Universe}}, in:
  \bibinfo{booktitle}{Bulletin of the American Astronomical Society}, p.
  \bibinfo{pages}{245}.
\newblock \href{http://arxiv.org/abs/1907.07558}{{\tt arXiv:1907.07558}}.
\bibitem[{{Meegan} et~al.(2009){Meegan}, {Lichti}, {Bhat}, {Bissaldi},
  {Briggs}, {Connaughton}, {Diehl}, {Fishman}, {Greiner}, {Hoover}, {van der
  Horst}, {von Kienlin}, {Kippen}, {Kouveliotou}, {McBreen}, {Paciesas},
  {Preece}, {Steinle}, {Wallace}, {Wilson} and
  {Wilson-Hodge}}]{2009ApJ...702..791M}
\bibinfo{author}{{Meegan}, C.}, \bibinfo{author}{{Lichti}, G.},
  \bibinfo{author}{{Bhat}, P.N.}, \bibinfo{author}{{Bissaldi}, E.},
  \bibinfo{author}{{Briggs}, M.S.}, \bibinfo{author}{{Connaughton}, V.},
  \bibinfo{author}{{Diehl}, R.}, \bibinfo{author}{{Fishman}, G.},
  \bibinfo{author}{{Greiner}, J.}, \bibinfo{author}{{Hoover}, A.S.},
  \bibinfo{author}{{van der Horst}, A.J.}, \bibinfo{author}{{von Kienlin}, A.},
  \bibinfo{author}{{Kippen}, R.M.}, \bibinfo{author}{{Kouveliotou}, C.},
  \bibinfo{author}{{McBreen}, S.}, \bibinfo{author}{{Paciesas}, W.S.},
  \bibinfo{author}{{Preece}, R.}, \bibinfo{author}{{Steinle}, H.},
  \bibinfo{author}{{Wallace}, M.S.}, \bibinfo{author}{{Wilson}, R.B.},
  \bibinfo{author}{{Wilson-Hodge}, C.}, \bibinfo{year}{2009}.
\newblock \bibinfo{title}{{The Fermi Gamma-ray Burst Monitor}}.
\newblock \bibinfo{journal}{\apj} \bibinfo{volume}{702},
  \bibinfo{pages}{791--804}.
\newblock \DOIprefix\doi{10.1088/0004-637X/702/1/791},
  \href{http://arxiv.org/abs/0908.0450}{{\tt arXiv:0908.0450}}.
\bibitem[{{Miller} et~al.(2019){Miller}, {Allen}, {Bellm}, {Bianco},
  {Blakeslee}, {Blum}, {Bolton}, {Briceno}, {Clarkson}, {Elias}, {Gezari},
  {Goodrich}, {Graham}, {Graham}, {Heathcote}, {Hsieh}, {Lotz}, {Matheson},
  {McSwain}, {Norman}, {Rector}, {Riddle}, {Ridgway}, {Saha}, {Street},
  {Soares-Santos}, {Skidmore}, {Stanghellini}, {Strolger}, {Thomas-Osip} and
  {Vivas}}]{2019BAAS...51g.154M}
\bibinfo{author}{{Miller}, B.}, \bibinfo{author}{{Allen}, L.},
  \bibinfo{author}{{Bellm}, E.}, \bibinfo{author}{{Bianco}, F.},
  \bibinfo{author}{{Blakeslee}, J.}, \bibinfo{author}{{Blum}, R.},
  \bibinfo{author}{{Bolton}, A.}, \bibinfo{author}{{Briceno}, C.},
  \bibinfo{author}{{Clarkson}, W.}, \bibinfo{author}{{Elias}, J.},
  \bibinfo{author}{{Gezari}, S.}, \bibinfo{author}{{Goodrich}, B.},
  \bibinfo{author}{{Graham}, M.}, \bibinfo{author}{{Graham}, M.},
  \bibinfo{author}{{Heathcote}, S.}, \bibinfo{author}{{Hsieh}, H.},
  \bibinfo{author}{{Lotz}, J.}, \bibinfo{author}{{Matheson}, T.},
  \bibinfo{author}{{McSwain}, M.V.}, \bibinfo{author}{{Norman}, D.},
  \bibinfo{author}{{Rector}, T.}, \bibinfo{author}{{Riddle}, R.},
  \bibinfo{author}{{Ridgway}, S.}, \bibinfo{author}{{Saha}, A.},
  \bibinfo{author}{{Street}, R.}, \bibinfo{author}{{Soares-Santos}, M.},
  \bibinfo{author}{{Skidmore}, W.}, \bibinfo{author}{{Stanghellini}, L.},
  \bibinfo{author}{{Strolger}, L.}, \bibinfo{author}{{Thomas-Osip}, J.},
  \bibinfo{author}{{Vivas}, K.}, \bibinfo{year}{2019}.
\newblock \bibinfo{title}{{Infrastructure and Strategies for Time Domain and
  MMA and Follow-Up}}, in: \bibinfo{booktitle}{Bulletin of the American
  Astronomical Society}, p. \bibinfo{pages}{154}.
\newblock \href{http://arxiv.org/abs/1908.11417}{{\tt arXiv:1908.11417}}.
\bibitem[{{Reitze} et~al.(2019){Reitze}, {LIGO Laboratory: California Institute
  of Technology}, {LIGO Laboratory: Massachusetts Institute of Technology},
  {LIGO Hanford Observatory} and {LIGO Livingston
  Observatory}}]{2019BAAS...51c.141R}
\bibinfo{author}{{Reitze}, D.}, \bibinfo{author}{{LIGO Laboratory: California
  Institute of Technology}}, \bibinfo{author}{{LIGO Laboratory: Massachusetts
  Institute of Technology}}, \bibinfo{author}{{LIGO Hanford Observatory}},
  \bibinfo{author}{{LIGO Livingston Observatory}}, \bibinfo{year}{2019}.
\newblock \bibinfo{title}{{The US Program in Ground-Based Gravitational Wave
  Science: Contribution from the LIGO Laboratory}}.
\newblock \bibinfo{journal}{\baas} \bibinfo{volume}{51}, \bibinfo{pages}{141}.
\newblock \href{http://arxiv.org/abs/1903.04615}{{\tt arXiv:1903.04615}}.
\bibitem[{Roberts et~al.(2021)Roberts, Burke, Benson, Lubelczyk, Bradley,
  Heckler and Hudiburg}]{roberts2021evaluation}
\bibinfo{author}{Roberts, C.J.}, \bibinfo{author}{Burke, J.C.},
  \bibinfo{author}{Benson, M.J.}, \bibinfo{author}{Lubelczyk, J.T.},
  \bibinfo{author}{Bradley, T.H.}, \bibinfo{author}{Heckler, G.W.},
  \bibinfo{author}{Hudiburg, J.J.}, \bibinfo{year}{2021}.
\newblock \bibinfo{title}{Evaluation of timely communications access methods
  using nasa space network}.
\newblock \bibinfo{journal}{Journal of Aerospace Information Systems} ,
  \bibinfo{pages}{1--14}\DOIprefix\doi{10.2514/1.I010897}.
\bibitem[{{Salgado} et~al.(2021){Salgado}, {Ibarra}, {Ehle}, {Gabriel},
  {Kretschmar}, {Kuulkers}, {Mer{\'\i}n}, {Ness}, {Salazar}, {Saxton},
  {Cecconi}, {Foster}, {Demleitner}, {Dempsey}, {Molinaro}, {S{\'a}nchez},
  {Taylor}, {D{\'\i}az Trigo}, {Fern{\'a}ndez}, {Kennea}, {Kettenis}, {Matt},
  {Osborne}, {de O{\~n}a Wilhelmi}, {Salbol}, {Sivakoff}, {Tao}, {Tohuvavohu},
  {Tibbetts} and {Workman}}]{2021ivoa.spec.0724S}
\bibinfo{author}{{Salgado}, J.}, \bibinfo{author}{{Ibarra}, A.},
  \bibinfo{author}{{Ehle}, M.}, \bibinfo{author}{{Gabriel}, C.},
  \bibinfo{author}{{Kretschmar}, P.}, \bibinfo{author}{{Kuulkers}, E.},
  \bibinfo{author}{{Mer{\'\i}n}, B.}, \bibinfo{author}{{Ness}, J.U.},
  \bibinfo{author}{{Salazar}, E.}, \bibinfo{author}{{Saxton}, R.},
  \bibinfo{author}{{Cecconi}, B.}, \bibinfo{author}{{Foster}, K.},
  \bibinfo{author}{{Demleitner}, M.}, \bibinfo{author}{{Dempsey}, J.},
  \bibinfo{author}{{Molinaro}, M.}, \bibinfo{author}{{S{\'a}nchez}, C.},
  \bibinfo{author}{{Taylor}, M.}, \bibinfo{author}{{D{\'\i}az Trigo}, M.},
  \bibinfo{author}{{Fern{\'a}ndez}, M.}, \bibinfo{author}{{Kennea}, J.},
  \bibinfo{author}{{Kettenis}, M.}, \bibinfo{author}{{Matt}, G.},
  \bibinfo{author}{{Osborne}, J.}, \bibinfo{author}{{de O{\~n}a Wilhelmi}, E.},
  \bibinfo{author}{{Salbol}, E.J.}, \bibinfo{author}{{Sivakoff}, G.},
  \bibinfo{author}{{Tao}, L.}, \bibinfo{author}{{Tohuvavohu}, A.},
  \bibinfo{author}{{Tibbetts}, M.}, \bibinfo{author}{{Workman}, B.},
  \bibinfo{year}{2021}.
\newblock \bibinfo{title}{{Observation Locator Table Access Protocol Version
  1.0}}.
\newblock \bibinfo{howpublished}{IVOA Recommendation 24 July 2021}.
\bibitem[{{Smith} et~al.(2013){Smith}, {Fox}, {Cowen}, {M{\'e}sz{\'a}ros},
  {Te{\v{s}}i{\'c}}, {Fixelle}, {Bartos}, {Sommers}, {Ashtekar}, {Jogesh Babu},
  {Barthelmy}, {Coutu}, {DeYoung}, {Falcone}, {Gao}, {Hashemi}, {Homeier},
  {M{\'a}rka}, {Owen} and {Taboada}}]{2013APh....45...56S}
\bibinfo{author}{{Smith}, M.W.E.}, \bibinfo{author}{{Fox}, D.B.},
  \bibinfo{author}{{Cowen}, D.F.}, \bibinfo{author}{{M{\'e}sz{\'a}ros}, P.},
  \bibinfo{author}{{Te{\v{s}}i{\'c}}, G.}, \bibinfo{author}{{Fixelle}, J.},
  \bibinfo{author}{{Bartos}, I.}, \bibinfo{author}{{Sommers}, P.},
  \bibinfo{author}{{Ashtekar}, A.}, \bibinfo{author}{{Jogesh Babu}, G.},
  \bibinfo{author}{{Barthelmy}, S.D.}, \bibinfo{author}{{Coutu}, S.},
  \bibinfo{author}{{DeYoung}, T.}, \bibinfo{author}{{Falcone}, A.D.},
  \bibinfo{author}{{Gao}, S.}, \bibinfo{author}{{Hashemi}, B.},
  \bibinfo{author}{{Homeier}, A.}, \bibinfo{author}{{M{\'a}rka}, S.},
  \bibinfo{author}{{Owen}, B.J.}, \bibinfo{author}{{Taboada}, I.},
  \bibinfo{year}{2013}.
\newblock \bibinfo{title}{{The Astrophysical Multimessenger Observatory Network
  (AMON)}}.
\newblock \bibinfo{journal}{Astroparticle Physics} \bibinfo{volume}{45},
  \bibinfo{pages}{56--70}.
\newblock \DOIprefix\doi{10.1016/j.astropartphys.2013.03.003},
  \href{http://arxiv.org/abs/1211.5602}{{\tt arXiv:1211.5602}}.
\bibitem[{{Stein} et~al.(2021){Stein}, {Velzen}, {Kowalski}, {Franckowiak},
  {Gezari}, {Miller-Jones}, {Frederick}, {Sfaradi}, {Bietenholz}, {Horesh},
  {Fender}, {Garrappa}, {Ahumada}, {Andreoni}, {Belicki}, {Bellm},
  {B{\"o}ttcher}, {Brinnel}, {Burruss}, {Cenko}, {Coughlin}, {Cunningham},
  {Drake}, {Farrar}, {Feeney}, {Foley}, {Gal-Yam}, {Golkhou}, {Goobar},
  {Graham}, {Hammerstein}, {Helou}, {Hung}, {Kasliwal}, {Kilpatrick}, {Kong},
  {Kupfer}, {Laher}, {Mahabal}, {Masci}, {Necker}, {Nordin}, {Perley},
  {Rigault}, {Reusch}, {Rodriguez}, {Rojas-Bravo}, {Rusholme}, {Shupe},
  {Singer}, {Sollerman}, {Soumagnac}, {Stern}, {Taggart}, {van Santen}, {Ward},
  {Woudt} and {Yao}}]{2021NatAs...5..510S}
\bibinfo{author}{{Stein}, R.}, \bibinfo{author}{{Velzen}, S.v.},
  \bibinfo{author}{{Kowalski}, M.}, \bibinfo{author}{{Franckowiak}, A.},
  \bibinfo{author}{{Gezari}, S.}, \bibinfo{author}{{Miller-Jones}, J.C.A.},
  \bibinfo{author}{{Frederick}, S.}, \bibinfo{author}{{Sfaradi}, I.},
  \bibinfo{author}{{Bietenholz}, M.F.}, \bibinfo{author}{{Horesh}, A.},
  \bibinfo{author}{{Fender}, R.}, \bibinfo{author}{{Garrappa}, S.},
  \bibinfo{author}{{Ahumada}, T.}, \bibinfo{author}{{Andreoni}, I.},
  \bibinfo{author}{{Belicki}, J.}, \bibinfo{author}{{Bellm}, E.C.},
  \bibinfo{author}{{B{\"o}ttcher}, M.}, \bibinfo{author}{{Brinnel}, V.},
  \bibinfo{author}{{Burruss}, R.}, \bibinfo{author}{{Cenko}, S.B.},
  \bibinfo{author}{{Coughlin}, M.W.}, \bibinfo{author}{{Cunningham}, V.},
  \bibinfo{author}{{Drake}, A.}, \bibinfo{author}{{Farrar}, G.R.},
  \bibinfo{author}{{Feeney}, M.}, \bibinfo{author}{{Foley}, R.J.},
  \bibinfo{author}{{Gal-Yam}, A.}, \bibinfo{author}{{Golkhou}, V.Z.},
  \bibinfo{author}{{Goobar}, A.}, \bibinfo{author}{{Graham}, M.J.},
  \bibinfo{author}{{Hammerstein}, E.}, \bibinfo{author}{{Helou}, G.},
  \bibinfo{author}{{Hung}, T.}, \bibinfo{author}{{Kasliwal}, M.M.},
  \bibinfo{author}{{Kilpatrick}, C.D.}, \bibinfo{author}{{Kong}, A.K.H.},
  \bibinfo{author}{{Kupfer}, T.}, \bibinfo{author}{{Laher}, R.R.},
  \bibinfo{author}{{Mahabal}, A.A.}, \bibinfo{author}{{Masci}, F.J.},
  \bibinfo{author}{{Necker}, J.}, \bibinfo{author}{{Nordin}, J.},
  \bibinfo{author}{{Perley}, D.A.}, \bibinfo{author}{{Rigault}, M.},
  \bibinfo{author}{{Reusch}, S.}, \bibinfo{author}{{Rodriguez}, H.},
  \bibinfo{author}{{Rojas-Bravo}, C.}, \bibinfo{author}{{Rusholme}, B.},
  \bibinfo{author}{{Shupe}, D.L.}, \bibinfo{author}{{Singer}, L.P.},
  \bibinfo{author}{{Sollerman}, J.}, \bibinfo{author}{{Soumagnac}, M.T.},
  \bibinfo{author}{{Stern}, D.}, \bibinfo{author}{{Taggart}, K.},
  \bibinfo{author}{{van Santen}, J.}, \bibinfo{author}{{Ward}, C.},
  \bibinfo{author}{{Woudt}, P.}, \bibinfo{author}{{Yao}, Y.},
  \bibinfo{year}{2021}.
\newblock \bibinfo{title}{{A tidal disruption event coincident with a
  high-energy neutrino}}.
\newblock \bibinfo{journal}{Nature Astronomy} \bibinfo{volume}{5},
  \bibinfo{pages}{510--518}.
\newblock \DOIprefix\doi{10.1038/s41550-020-01295-8},
  \href{http://arxiv.org/abs/2005.05340}{{\tt arXiv:2005.05340}}.
\bibitem[{{Tohuvavohu} et~al.(2020){Tohuvavohu}, {Kennea}, {DeLaunay},
  {Palmer}, {Cenko} and {Barthelmy}}]{2020ApJ...900...35T}
\bibinfo{author}{{Tohuvavohu}, A.}, \bibinfo{author}{{Kennea}, J.A.},
  \bibinfo{author}{{DeLaunay}, J.}, \bibinfo{author}{{Palmer}, D.M.},
  \bibinfo{author}{{Cenko}, S.B.}, \bibinfo{author}{{Barthelmy}, S.},
  \bibinfo{year}{2020}.
\newblock \bibinfo{title}{{Gamma-Ray Urgent Archiver for Novel Opportunities
  (GUANO): Swift/BAT Event Data Dumps on Demand to Enable Sensitive
  Subthreshold GRB Searches}}.
\newblock \bibinfo{journal}{\apj} \bibinfo{volume}{900}, \bibinfo{pages}{35}.
\newblock \DOIprefix\doi{10.3847/1538-4357/aba94f},
  \href{http://arxiv.org/abs/2005.01751}{{\tt arXiv:2005.01751}}.
\bibitem[{{Winkler} et~al.(2003){Winkler}, {Courvoisier}, {Di Cocco},
  {Gehrels}, {Gim{\'e}nez}, {Grebenev}, {Hermsen}, {Mas-Hesse}, {Lebrun},
  {Lund}, {Palumbo}, {Paul}, {Roques}, {Schnopper}, {Sch{\"o}nfelder},
  {Sunyaev}, {Teegarden}, {Ubertini}, {Vedrenne} and
  {Dean}}]{2003A&A...411L...1W}
\bibinfo{author}{{Winkler}, C.}, \bibinfo{author}{{Courvoisier}, T.J.L.},
  \bibinfo{author}{{Di Cocco}, G.}, \bibinfo{author}{{Gehrels}, N.},
  \bibinfo{author}{{Gim{\'e}nez}, A.}, \bibinfo{author}{{Grebenev}, S.},
  \bibinfo{author}{{Hermsen}, W.}, \bibinfo{author}{{Mas-Hesse}, J.M.},
  \bibinfo{author}{{Lebrun}, F.}, \bibinfo{author}{{Lund}, N.},
  \bibinfo{author}{{Palumbo}, G.G.C.}, \bibinfo{author}{{Paul}, J.},
  \bibinfo{author}{{Roques}, J.P.}, \bibinfo{author}{{Schnopper}, H.},
  \bibinfo{author}{{Sch{\"o}nfelder}, V.}, \bibinfo{author}{{Sunyaev}, R.},
  \bibinfo{author}{{Teegarden}, B.}, \bibinfo{author}{{Ubertini}, P.},
  \bibinfo{author}{{Vedrenne}, G.}, \bibinfo{author}{{Dean}, A.J.},
  \bibinfo{year}{2003}.
\newblock \bibinfo{title}{{The INTEGRAL mission}}.
\newblock \bibinfo{journal}{\aap} \bibinfo{volume}{411},
  \bibinfo{pages}{L1--L6}.
\newblock \DOIprefix\doi{10.1051/0004-6361:20031288}.

\end{thebibliography}

\end{document}